\documentclass[runningheads]{llncs}
\usepackage[T1]{fontenc}
\usepackage{graphicx}
\usepackage{array}
\usepackage{mdframed}
\usepackage{color}
\usepackage{amssymb}
\usepackage{caption}
\usepackage{subcaption}
\usepackage{multirow, makecell}
\usepackage{hhline}
\newcolumntype{C}[1]{>{\centering\let\newline\\\arraybackslash\hspace{0pt}}m{#1}}
\newcolumntype{L}[1]{>{\raggedright\let\newline\\\arraybackslash\hspace{0pt}}m{#1}}
\newcolumntype{R}[1]{>{\raggedleft\let\newline\\\arraybackslash\hspace{0pt}}m{#1}}

\begin{document}
\title{mdTLS: How to make middlebox-aware TLS more efficient?}

\author{Taehyun Ahn\orcidID{0009-0007-2339-286X} \and
Jiwon Kwak\orcidID{0009-0008-1560-7442} \and
Seungjoo Kim\thanks{Corresponding Author}\orcidID{0000-0002-2157-0403}}

\authorrunning{Ahn et al.}

\institute{School of Cybersecurity, Korea University, Seoul 02841, South Korea \\
\email{\{thyun\_ahn,jwkwak4031,skim71\}@korea.ac.kr}}

\maketitle

\begin{abstract}
Recently, many organizations have been installing middleboxes in their networks in large numbers to provide various services to their customers. Although middleboxes have the advantage of not being dependent on specific hardware and being able to provide a variety of services, they can become a new attack target for hackers. Therefore, many researchers have proposed security-enchanced TLS protocols, but their results have some limitations. In this paper, we proposed a middlebox-delegated TLS (mdTLS) protocol that not only achieves the same security level but also requires relatively less computation compared to recent research results. mdTLS is a TLS protocol designed based on the proxy signature scheme, which requires about 39\% less computation than middlebox-aware TLS (maTLS), which is the best in security and performance among existing research results. \textcolor{black}{In order to substantiate the enhanced security of mdTLS, we conducted a formal verification using the Tamarin.} Our verification demonstrates that mdTLS not only satisfies the security properties set forth by maTLS but also complies with the essential security properties required for proxy signature scheme.\footnote{All of the formal models and lemmas are open to the public through the following url https://github.com/HackProof/mdTLS}

%% 최근 많은 기업에서 고객에게 다양한 서비스를 제공하기 위해 네트워크에 미들박스를 대량으로 설치하고 있습니다. 미들박스는 특정 하드웨어에 종속되지 않고 다양한 서비스를 제공할 수 있다는 장점이 있지만, 해커에게는 새로운 공격 대상이 될 수 있습니다. 따라서 많은 연구자들이 미들박스를 고려하여 네트워크의 보안을 강화하기 위한 개선된 TLS 프로토콜을 제안하고 있지만, 그 결과에는 몇 가지 한계가 있습니다. 본 논문에서는 최근 연구 결과와 비교해 동일한 보안 수준을 달성할 뿐만 아니라 상대적으로 연산량이 적은 미들박스 위임형 TLS(mdTLS) 프로토콜을 제안합니다. mdTLS는 프록시 서명 기법을 기반으로 설계된 TLS 프로토콜로, 기존 연구 결과 중 보안 및 성능 측면에서 가장 우수한 미들박스 인식형 TLS(maTLS) 대비 약 39\% 적은 연산량을 필요로 합니다. 보안을 위해 성능을 희생할 필요가 없다는 것을 증명하기 위해 연구진은 mdTLS가 maTLS가 입증한 보안 목표뿐만 아니라 프록시 서명 기술이 충족해야 하는 보안 속성도 충족한다는 것을 공식적으로 검증했습니다. mdTLS의 강화된 보안이 성능 저하를 수반하지 않는다는 것을 입증하기 위해 우리는 Tamarin 도구를 이용하여 공식적인 검증을 수행하여 mdTLS가 maTLS가 설정한 보안 목표를 충족할 뿐만 아니라 프록시 서명 기술에 요구되는 필수 보안 속성도 준수한다는 것을 입증했습니다.


\end{abstract}

\section{Introduction}
The advent of the COVID-19 pandemic has instigated substantial transformations in the business landscape. Notably, a significant proportion of enterprises have transitioned from conventional in-office working arrangements to facilitating remote work options for their workforce. Concurrently, the pandemic has spurred innovative shifts in operational methodologies, exemplified by the substitution of face-to-face business procedures, historically reliant on in-person meetings, with video conferencing solutions. As a result of these shifts, there has been a discernible escalation in network traffic, with notable statistics from the Telegraph indicating a remarkable 47\% surge in internet traffic between 2019 and 2020 \cite{ref_url1}.\\
%%코로나19 팬데믹의 도래로 비즈니스 환경이 크게 변화하고 있습니다. 특히 상당수의 기업이 기존의 사무실 내 근무 방식에서 직원들을 위한 원격 근무 옵션으로 전환했습니다. 동시에 팬데믹으로 인해 과거에는 대면 회의에 의존했던 대면 비즈니스 절차가 화상 회의 솔루션으로 대체되는 등 운영 방법론의 혁신적인 변화가 촉진되었습니다. 이러한 변화의 결과로 네트워크 트래픽이 눈에 띄게 증가했으며, 텔레그래프의 주목할 만한 통계에 따르면 2019년과 2020년 사이에 인터넷 트래픽이 47% 급증한 것으로 나타났습니다.
Especially during the COVID-19 pandemic, the security of confidential information of various companies and individuals has been emphasized as most social activities, including business, are conducted remotely over the network. Among the most prominent and widely adopted technologies addressing network security concerns during this period is HTTPS (HyperText Transfer Protocol Secure) \cite{ref_url13}. 
%% 코로나19 팬데믹의 도래로 비즈니스 환경이 크게 변화하고 있습니다. 특히 상당수의 기업이 기존의 사무실 내 근무 방식에서 직원들을 위한 원격 근무 옵션으로 전환했습니다. 동시에 팬데믹으로 인해 과거에는 대면 회의에 의존했던 대면 비즈니스 절차가 화상 회의 솔루션으로 대체되는 등 운영 방법론의 혁신적인 변화가 촉진되었습니다. 이러한 변화의 결과로 네트워크 트래픽이 눈에 띄게 증가했으며, 텔레그래프의 주목할 만한 통계에 따르면 2019년과 2020년 사이에 인터넷 트래픽이 47% 급증한 것으로 나타났습니다. 특히 코로나 사태를 겪으면서, 비즈니스를 포함한 대부분의 사회 활동들이 네트워크를 통한 원격으로 수행되면서 다양한 회사 및 개인의 기밀정보에 대한 보안이 강조되고 있다. 가장 대표적인 네트워크 보안 기술은 HTTPS이다. 

HTTPS represents a communication protocol that integrates the HTTP (HyperText Transfer Protocol) \cite{ref_url14} to the TLS (Transport Layer Security) protocol \cite{ref_url15}, with the overarching objective of ensuring the confidentiality and integrity of data transmitted over networks. \textcolor{black}{This protocol finds utility not only in desktops but extends its application domain to encompass a diverse array of embedded devices, including IoT (Internet of Things) devices.} HTTPS offers several fundamental security attributes, including the following:
\begin{itemize}
    \item \textit{Encryption:} It serves as a pivotal mechanism within HTTPS, facilitating the obfuscation of sensitive information by encoding the data exchanged between communicating entities. Commonly employed encryption algorithms encompass symmetric key algorithms like Advanced Encryption Standard (AES) \cite{ref_article9}.
    \item \textit{Authentication:} It constitutes an integral component of HTTPS, operating to ascertain the identity of entities by utilizing digital certificates.
    \item \textit{Integrity:} It is another crucial facet of HTTPS, operating as a mechanism to detect unauthorized tampering or forgery of messages. Conventional algorithms used to maintain message integrity involve the implementation of Message Authentication Codes (MACs), such as the Secure Hash Algorithm (SHA) \cite{ref_article10}, to uphold the veracity and unaltered state of a network connection.
\end{itemize}
%%HTTPS는 네트워크를 통해 전송되는 데이터의 기밀성과 무결성을 보장하기 위해 HTTP와 TLS 프로토콜을 통합한 통신 프로토콜입니다. 해당 프로토콜은 일반 PC뿐만 아니라 IoT 디바이스 등 다양한 임베디드 디바이스에서도 사용되고 있으며 다음과 같은 기본적인 보안 성질을 제공한다.
%%- 암호화: 통신하는 개체 간에 교환되는 데이터를 암호화하여 데이터 안의 기밀 정보를 숨기는 메커니즘으로, 암호화 시 활용되는 대표적인 알고리즘으로 AES 와 같은 대칭키 알고리즘 등이 존재한다.
%%- 인증: 인증서를 사용하여 사용자의 신원을 식별하는 메커니즘이다.
%%- 무결성: 메시지 변조 및 위조를 탐지하는 메커니즘으로, 대표적인 알고리즘으로 연결의 신뢰성을 보장하며 보안 해시 알고리즘(SHA: Secure Hash Algorithm)과 같은 메시지 인증 코드(MAC: Message Authentication Code) 등이 존재한다.
According to the Google transparency report, there has been a consistent increase in the loading speed of HTTPS pages in the chrome browser since 2014 \cite{ref_url2}. Moreover, among the top 100 non-Google websites on the internet, which collectively constitute approximately 25\% of global website traffic, 96 websites have embraced HTTPS, with 90 of them making HTTPS their default protocol. Additionally, according to Gartner's article \cite{ref_url3}, edge computing technology is anticipated to evolve into a core IT technology. This technology facilitates the secure communications of data collected through embedded systems deployed across various domains, relying on TLS protocols. Consequently, TLS communication is expected to assume an increasingly pivotal role. However, the robust encryption mechanisms employed by TLS to protect data can also be exploited by attackers to hide malware within network traffic, thereby evading detection by conventional security measures. In fact, according to research by Cisco and Sophos, TLS is vulnerable to detecting malicious traffic, and the number of such cases continues to increase \cite{ref_url18,ref_url19}. As a result, TLS cannot be considered a complete solution against cybersecurity threats.\\
%However, it is essential to note that the TLS protocol, while paramount for ensuring data security, does not inherently provide supplementary functionalities such as malicious traffic detection, caching, or content optimization \cite{ref_url4,ref_proc1}.
%% Google 투명성 보고서에 따르면 Chrome 브라우저에서 HTTPS 페이지의 로딩 속도는 2014년 이후 꾸준히 증가하고 있습니다. 또한 전 세계 웹사이트 트래픽의 약 25\%를 차지하는 인터넷 상위 100대 비구글 사이트 중 96개 웹사이트가 HTTPS를 지원하며 90개 웹사이트가 HTTPS를 기본값으로 설정하고 있습니다. 또한 가트너의 '2019년 10대 전략 기술 트렌드'에 따르면, 다양한 분야에 설치된 임베디드 시스템을 통해 수집된 정보가 TLS 통신을 기반으로 사용자에게 실시간으로 안전하게 제공될 수 있는 엣지 컴퓨팅 기술은 점차 핵심 IT 기술로 성장할 것으로 전망됩니다. 이에 따라 TLS 통신의 중요성은 더욱 높아질 것으로 예상됩니다. 하지만 TLS 프로토콜은 악성 트래픽 탐지, 캐싱, 콘텐츠 최적화 등의 부가 기능을 제공하지 않기 때문에 사이버공격으로 부터 완벽히 안전하다고 볼 수 없다.
 For this reason, numerous organizations have deployed specialized middleboxes with distinct functionalities designed to enhance security for their clients, such as firewall and intrusion detection \cite{ref_proc1}. For instance, some companies have integrated Transport Layer Security Inspection (TLSI) \cite{ref_url5} capabilities into select middleboxes to identify and intercept malicious traffic attempting to infiltrate their internal networks. TLSI represents a technology devised to thwart unauthorized actions perpetrated by hackers on encrypted network traffic, and this technology is being actively leveraged by numerous entities, including industry giants such as Microsoft \cite{ref_url6}.\\
 However, according to a survey conducted in the United States, more than 70\% of employees still believe that hackers can exploit middleboxes. Also, 50\% of the respondents answered that their personal information could be infringed by exploiting vulnerabilities in the middleboxes \cite{ref_article1}. Ironically, middleboxes, initially installed to fortify data security within TLS communications, have emerged as potential targets for cyberattacks. Consequently, safeguarding data transmitted over TLS communications necessitates a holistic approach considering network components, such as middleboxes, from the inception of communication channel construction. \textcolor{black}{This approach goes beyond simply installing security-hardened components into an existing network.} \\
%%이러한 이유로 많은 조직에서는 고객에게 안전하게 서비스를 제공하기 위해 특정 기능이 내장된 미들박스를 설치했습니다. 예를 들어, 기업들은 사내 망으로 유입되는 악성 트래픽을 탐지하기 위해 TLSI(전송 계층 보안 검사) 기능을 특정 미들박스에 내장하고 있습니다. TLSI는 해커가 암호화된 트래픽에서 허가되지 않은 행위를 하지 못하도록 보호하는 기술로, Microsoft를 비롯한 많은 기업이 이 기술을 활용하고 있습니다. 하지만 미국에서 실시한 설문조사에 따르면 70 % 이상의 직원은 여전히 해커가 미들박스를 악용할 수 있다고 응답했습니다. 또한 응답자의 50 %는 미들박스의 취약점을 악용하여 개인정보가 침해될 수 있다고 답했습니다. 역설적이게도 TLS 통신 프로토콜을 통해 전송되는 데이터에 대한 보안을 향상시키기 위해, 설치된 미들박스들이 오히려 사이버공격의 표적이 될 수 있는 상황이다. 이에 따라 TLS 통신 시 전송되는 데이터를 보호하기 위해서는 보안기능이 탑재된 구성 요소를 단순히 네트워크에 설치하는 것이 아니라, 통신 채널을 구축하는 초기부터 미들박스 등의 네트워크 구성요소가 고려되는 것이 필요하다.

As a consequence, numerous researchers have proposed a range of TLS extension protocols to enhance security during communication via the TLS protocol. \textcolor{black}{However, prior research endeavors, driven primarily by a pursuit of security, have inadvertently encountered performance-related challenges.} In this study, we will introduce the mdTLS protocol, which is meticulously designed based on proxy signature scheme. \textcolor{black}{The mdTLS is subject to comparative evaluation against maTLS \cite{ref_proc8}, widely recognized as the most exemplary among prior researches in terms of both security and performance.} First, we investigated the amount of arithmetic operations that must be performed for each designed protocol to compare the performance of the mdTLS and maTLS protocols. We then formally verified that the mdTLS satisfies not only the security properties verified in maTLS, but also three other security properties related to proxy signature scheme. To ensure methodological consistency in our experimental setup, we employed the Tamarin \cite{ref_lncs2,ref_proc10,ref_url7}, utilized in prior maTLS research, during the security analysis.\\
%% 그 결과, 많은 연구자들이 TLS 프로토콜을 통해 통신할 때 보안을 강화하기 위한 다양한 TLS 확장 프로토콜을 제안했습니다. 그러나 기존 연구들은 너무 보안에 초점을 맞춰 성능 문제를 야기했다. 이에 따라 우리는 성능을 향상시키기 위해 프록시 서명을 기반으로 mdTLS 프로토콜을 설계하였습니다. 이후 우리는 기존에 제안되었된 다양한 TLS 확장 프로토콜 중 보안 및 성능 수준에서 가장 우수한 maTLS와 우리가 설계한 mdTLS를 보안 및 성능 2가지 측면에서 비교하였습니다. 먼저 우리는 mdTLS 프로토콜과 maTLS 프로토콜의 성능을 비교하기 위해 설계된 프로토콜 별 수행해야 하는 사칙 연산량을 조사했습니다. 이후 우리는 mdTLS 프로토콜이 maTLS에서 검증된 보안 성질뿐만 아니라 프록시 서명과 관련된 세 가지 다른 보안 성질도 충족한다는 것을 정형 검증하였습니다. 이때 우리는 동일한 실험 환경을 구축하기 위해 정형검증 수행 시, maTLS 연구에서 활용되었던 Tamarin 도구를 활용하였습니다.
The remainder of the paper is organized as follows. First, we analyzed the strengths and weaknesses of related works (Section 2). Next, we introduced our mdTLS protocol (Section 3). After that, we compare the performance between maTLS and mdTLS (Section 4). In Section 5, we verified our protocol using Tamarin (Section 5). We showed that the performance can be further improved when the Schnorr digital signature is used in the protocol (Section 6). Finally, we present our concluding remarks (Section 7). \newpage

\section{Related works}
Many researches have been conducted to improve TLS protocol. They are categorized into two types. One is the TLS-encryption extension-based approach. Their research is to improve the mechanism itself inside the protocol. The other one is the Trusted Execution Environment (TEE) based approach. Their research is to improve the protocol by using specific hardware.
\subsection{TEE based approaches}
A typical example of the Trusted Execution Environment (TEE) based approach is SGX-Box \cite{ref_proc2}. It utilized the remote attestation of Intel SGX. The server performs remote attestation to verify the integrity of the SGX-Box module in middleboxes. If remote attestation succeeds, they create a secure channel to prevent sensitive information from leaking between them. However, it is limited in that it is too dependent on its specific hardware (Intel SGX). Besides SGX-Box, there are many researches such as STYX \cite{ref_proc3}, EndBox \cite{ref_proc4}, and ShieldBox \cite{ref_proc5}. However, they also had the same limitations mentioned above.
\subsection{TLS-extension based approaches}
A typical example of the TLS-extension approach is SplitTLS \cite{ref_proc6}. In SplitTLS, middleboxes act as servers and clients at the same time. This feature gives them too many privileges. It can cause some security incidents. For example, middleboxes such as CDN service providers could receive the secret key to act as a server. It accidentally exposes the secret key during the key-exchange phase. The worst thing is that when the middleboxes become compromised, malicious users (attackers) could abuse their privileges. Unlike SplitTLS, mcTLS \cite{ref_article2} provides the least privilege to middleboxes. Middleboxes can read or write the TLS payload by obtaining MAC key pair from each endpoint. For example, they can only read the TLS packets when they get a unique key for reading. The advantage of mcTLS is that it does not force middleboxes to create or install further objects. Since the mcTLS uses only one key when creating a session, it is considered insecure. In the performance view, it has a limitation in that additional latency occurs when establishing the first connection. Furthermore, it does not follow TLS standards. David Naylor, who had proposed mcTLS, proposed an extended version of mcTLS called mbTLS \cite{ref_proc7}. mbTLS was created to improve compatibility with TLS standards. mbTLS establishes two types of sessions. One is the mbTLS session, and the other is the standard TLS session. If one of the endpoints does not use mbTLS, then traditional TLS sessions are activated. Overall, mbTLS offers improvements over mcTLS, which causes latency when adding a secondary session. maTLS \cite{ref_proc8} is another extended protocol to address security issues in SplitTLS. It treats middleboxes as equivalent entities to the server and includes them in the TLS session. As the server’s certificate, middleboxes’ certificates are issued by the Certificate Authority (CA), and by introducing the Middlebox Transparency (MT) log server, the middleboxes certificate contains a Signed Certificate Timestamp (SCT) \cite{ref_url9,ref_url16}. This guarantees middleboxes’ audition and improves the reliability of the middleboxes’ certificates. Also, unlike SplitTLS, this procedure shows middleboxes can create their own official certificates without using custom root certificates or server certificates. However, these security elements entail performance issues. To make every session in each section, maTLS handshakes are essential between every entity. This is why maTLS’s initial handshake takes more time than the original version of TLS.

\section{mdTLS: middlebox-delegated TLS protocol with proxy signature scheme}
In this section, we described the mdTLS protocol. At first, we defined the adversary model and security goals related to the mdTLS. After that, we described each phase in the protocol in detail.

\subsection{Adversary model}
We considered the attacker’s capability under the Dolev-Yao model \cite{ref_article3}. Attackers can obtain and analyze messages in the network. Furthermore, they can get public keys. They aim to obtain certificates, perform an impersonation attack via forged certificates, and reveal secret keys.

\subsection{Security goal}
TLS currently provides the following properties in multi-party cases. Among them, we define "secure" for mdTLS by extending three security properties to cover the "delegation" concept.
\subsubsection{Authentication:}
The notion of authentication was defined as that every entity must be able to verify whether they are talking to the "right person". This goal was divided into two sub-goals. First, each entity(client or server) can verify whether the other endpoint is operated by the expected middleboxes. It is called \textit{entity authentication}. 
Second, If a session between two endpoints consists of an ordered set of middleboxes $MB_1$ ... $MB_{n-1}$, then any data received by $MB_j$ must be a prefix of the data sent by $MB_{j-1}$ or $MB_{j+1}$, where $1<j<n-1$. It is called \textit{data authentication}. 
We refined \textit{entity authentication} into two security goals. First, the client ensures the delegated middleboxes by verifying the warrant in signature. It is called \textit{verifiability}. Second, each middlebox can be identified as an appropriately delegated middlebox by checking its public key from the proxy signature. It is called \textit{strong-identifiability}.

\newpage

\subsubsection{Secrecy:}
The notion of secrecy can be defined as that adversaries should learn nothing more from observing ciphertext in network connections. This goal is divided into two sub-goals. First, each mdTLS segment sent from entities should be encrypted with a strong ciphersuite. It is called \textit{segment secrecy}. Second, each segment should have its own security parameters, such as a unique session key, to prevent the data from being reused. It is called \textit{individual secrecy}.

\subsubsection{Integrity:}
The notion of integrity means that only authorized or delegated entities can make or modify messages under their permissions. This goal is divided into two sub-goals. First, the entity can confirm which middleboxes have made each modification to the message. It is called \textit{modification accountability}. Second, endpoints can determine the list and order of middleboxes that messages pass through. It is called \textit{path integrity}.
In mdTLS, we defined one security goal additionally. Delegated middleboxes can generate valid signatures. It means, in converse, undelegated entities cannot modify messages because they cannot generate and verify the signatures. Hence, it is called \textit{strong-unforgeability}.

\subsection{Overview of mdTLS protocol}
The mdTLS applies a proxy signature scheme based on the \textit{partial delegation with warrant} \cite{ref_article5,ref_lncs1,ref_article4} to improve performance while having the same security level as maTLS. \newline
Proxy signature scheme \cite{ref_article4} is a technique in which a proxy signer electronically signs on behalf of the original signer. When the original signer is temporarily absent, a proxy signer receives signature authority from the original signer and performs the proxy signing. This signing authority delegation technique can be used in various distributed systems, such as edge computing. There are four types of delegation in the proxy signature scheme: \textit{full delegation}, \textit{partial delegation}, \textit{delegation by warrant}, and \textit{partial delegation with warrant} \cite{ref_lncs1,ref_article4}.
 \begin{itemize}
     \item \textit{Full delegation:} The proxy signer uses the original signer's secret key to generate the proxy signature.
     \item \textit{Partial delegation:} This method generates a proxy signing key using the secret keys of both the original and the proxy signers. The advantage is that it can prevent the original signer from arbitrarily proxy signing, but there is no way to revoke or limit proxy signing authority.
     \item \textit{Delegation by warrant:} This method uses a warrant that specifies the proxy delegation period and message space to limit proxy signing authority. It can compensate for the shortcomings of partial delegation, but performance in verification deteriorates because the verifier must additionally verify the warrant when verifying the proxy signature.
     \item \textit{Partial delegation with warrant:} Kim et al. \cite{ref_lncs1} first introduced this type of delegation. This method utilizes the advantages of both \textit{partial delegation} and \textit{delegation by warrant}. Proxy signing authority can be restricted or revoked through a warrant. Additionally, since this method only verifies the proxy signature, the verification efficiency can be improved.
 \end{itemize}

The details of the mdTLS are shown in Figure 1, 2. For reader’s convenience, notation definitions are listed in Table 1. mdTLS is divided into 3 phases. 
\begin{itemize}
     \item \textit{Generating certificates phase:} Before negotiation, server certificates are generated.
     \item \textit{Handshake phase:} Negotiation between two endpoints on a network – such as a client and a server – to establish the details of their connection. During handshake, ECDH and ECDSA \cite{ref_article11,ref_article12} are used in key exchange and digital signature, respectively.
     \item \textit{Record phase:} Data communications are encrypted between the two entities.
 \end{itemize}

The following statements below Table 1 are detailed sequences in which each entity establishes a secure communication channel based on the mdTLS.
 
\begin{table}[h!]
\caption{Notations in mdTLS}\label{tab2}
\centering
\begin{tabular}{|C{15mm}|C{25mm}|L{75mm}|}
\hline
 & \multicolumn{1}{c|}{\textbf{Notation}} & \multicolumn{1}{c|}{\textbf{Meaning}}             \\
\hline
    \multirow{4}{*}{\textbf{Entities}}   & $C$        & Client    \\
                        & $S$        & Server    \\
                        & $MB_i$     & i-th middlebox ($0 < i < n$)    \\
                        & $e_i$     & i-th entity ($e_0$: client, $e_n$: server)    \\
\hline
    \textbf{ECDH} & $(d^{ex}_{e_i}, Q^{ex}_{e_i})$  &  $e_i$'s ECDH key pairs   \\
\hline
    \multirow{9}{*}{\textbf{ECDSA}} & $p$  &  A prime number   \\
     & $E$  &  An elliptic curve on $\mathbb{F}_p$   \\
     & $q$  &  A field size (prime number)   \\
     & $G$  &  A base point on $E$ having prime order $q$   \\
     & $d_{e_i}$  & A secret key with $0 < d{e_i} < q$   \\
     & $Q_{e_i}$  & A public key with $d_{e_i} \cdot G$ on $E$   \\
     & $H$  & Cryptographic hash function ($\{0,1\}^{*} \rightarrow \mathbb{F}_q$)   \\
     & $S^H(d_{e_i}, m)$  & Sign message $m$ with secret key $d_{e_i}$ using $H$   \\
     & $V^H(Q_{e_i}, m, \sigma)$  & Verify signature $\sigma$ generated by $S^H(d_{e_i}, m)$   \\
\hline
    \multirowcell{2}{\textbf{Proxy-} \\ \textbf{signature}} & $PS(skp, m)$  & Proxy signing the message $m$ with proxy signing key $skp$   \\
        & $PV(Q_{e_i}, m, \sigma_p)$  & Proxy verification for proxy signature $\sigma_p$, with $Q_{e_i}$  \\
\hline
\end{tabular}
\end{table}

\vspace{-0.5cm}

\subsubsection{Phase 0. Generating certificates}
\begin{enumerate}
    \item Server sends Certificate Signing Request (CSR) to Certificate Authority (CA).
    \item CA verifies CSR, creates pre-certificates, and submits to the Certificate Transparency (CT) log server to get SCTs \cite{ref_url9}.
    \item After the CT log server adds pre-certificates to the logs, it returns SCTs to CA. Due to the Certificate Transparency policy \cite{ref_url9,ref_url16}, at least 2 SCTs from different CT log servers are required for certificates.
    \item Using the X.509 v3 \cite{ref_url17} extension, CA attaches SCTs to the certificate and issues the certificate to the server.
\end{enumerate}

\subsubsection{Phase 1. Handshake}
\vspace{0.3cm}

% \begin{itemize}
%     \item \textbf{Notations}
% \begin{itemize}
%     \item Entities
%         \begin{itemize}
%             \item $C$: Client
%             \item $S$: Server
%             \item $MB_i$: i-th Middlebox ($0 < i < n$)
%             \item $e_i$: i-th Entity ($e_0: C$, $e_n: S$) 
%         \end{itemize}
%     \item ECDH key pair
%         \begin{itemize}
%             \item $(d^{ex}_{e_i}, Q^{ex}_{e_i})$: $e_i$'s ECDH key pairs 
%         \end{itemize}
%     \item ECDSA
%         \begin{itemize}
%             \item $p$: A prime number
%             \item $E$: An elliptic curve on $F_p$
%             \item $q$: A field size (prime number)
%             \item $G$: A base point on $E$ having prime order $q$
%             \item $d_{e_i}$: A secret key with $0 < d{e_i} < q$
%             \item $Q_{e_i}$: A public key with $d_{e_i} \cdot G$ on $E$
%             \item $H$ : $\{0,1\}^{*} \rightarrow F_{q}$
%             \item $S^H(d_{e_i}, m)$ : Sign message $m$ with secret key $d_{e_i}$ using Hash $H$
%             \item $V^H(Q_{e_i}, m, \sigma)$ : Verify signature $\sigma$ generated by $S^H(d_{e_i}, m)$ 
%         \end{itemize}
%     \item Proxy signature
%         \begin{itemize}
%             \item $PS(skp, m)$ : Proxy sign message $m$ with proxy signing key $skp$
%             \item $PV(Q_{e_i}, m, \sigma_p)$ : Proxy verification for proxy signature $\sigma_p$, with $Q_{e_i}$
%         \end{itemize}
% \end{itemize}
% \end{itemize}

\begin{enumerate}
\item Client generates ECDH key pair, and the public key $Q^{ex}_{C}$ will be sent by \textsf{ClientHello} message.
\item Middleboxes attach their two types of keys to the \textsf{ClientHello} message. One is ECDH public key, $Q^{ex}_{MB_{i}}$, and the other is ECDSA public key, $Q_{MB_i}$, which will be used in the proxy signature scheme. 
\item Server, the original signer, also creates its ECDH and ECDSA key pairs as middleboxes. When the server receives a \textsf{ClientHello} message, it operates the designation process to delegate middleboxes as proxy signers. Outputs of this process are called signed delegations $\sigma_{d}\__{MB_{i}}$. For delegation, the server has to sign the hash value of the delegation message. This message consists of $Q_{S}$, the identity of proxy signer $ID_{MB_{i}}$, $Q_{MB_{i}}$, and a warrant $\omega$ containing the message space and delegation period.
In addition, $0$ is prepended to represent that it is for the proxy signature scheme. $\sigma_{d}\__{MB_{i}}$ can be represented as $(x_{Y_d}, s_d)$ according to ECDSA form. Signed delegations will be sent by \textsf{ServerHello} message with $ Q^{ex}_{S}$.
\begin{itemize}        
    \item $\sigma_{d}\__{MB_{i}} \leftarrow S^H(d_{S},0||Q_{S}||ID_{MB_{i}}||Q_{MB_{i}}||\omega)$
    \begin{itemize}
        \item random value $y_d$ ( $0 < y_d < q$)
        \item $Y_d \leftarrow y_d \cdot G$
        \item $x_{Y_d} \leftarrow $ x-coordinate of $Y_d$
        \item $c \leftarrow H(m_d)$ $(m_d=0||Q_{S}||ID_{MB_{i}}||Q_{MB_{i}}||\omega)$
        \item $s_d \leftarrow (c + d_{S} \cdot x_{Y_d}) \cdot y_d^{-1}$ mod $q$
        \item $\therefore \sigma_{d}\__{MB_{i}} = (x_{Y_d}, s_d) = $ signed delegations 
    \end{itemize}
\end{itemize}

\begin{figure}[t!]
\includegraphics[width=\textwidth]{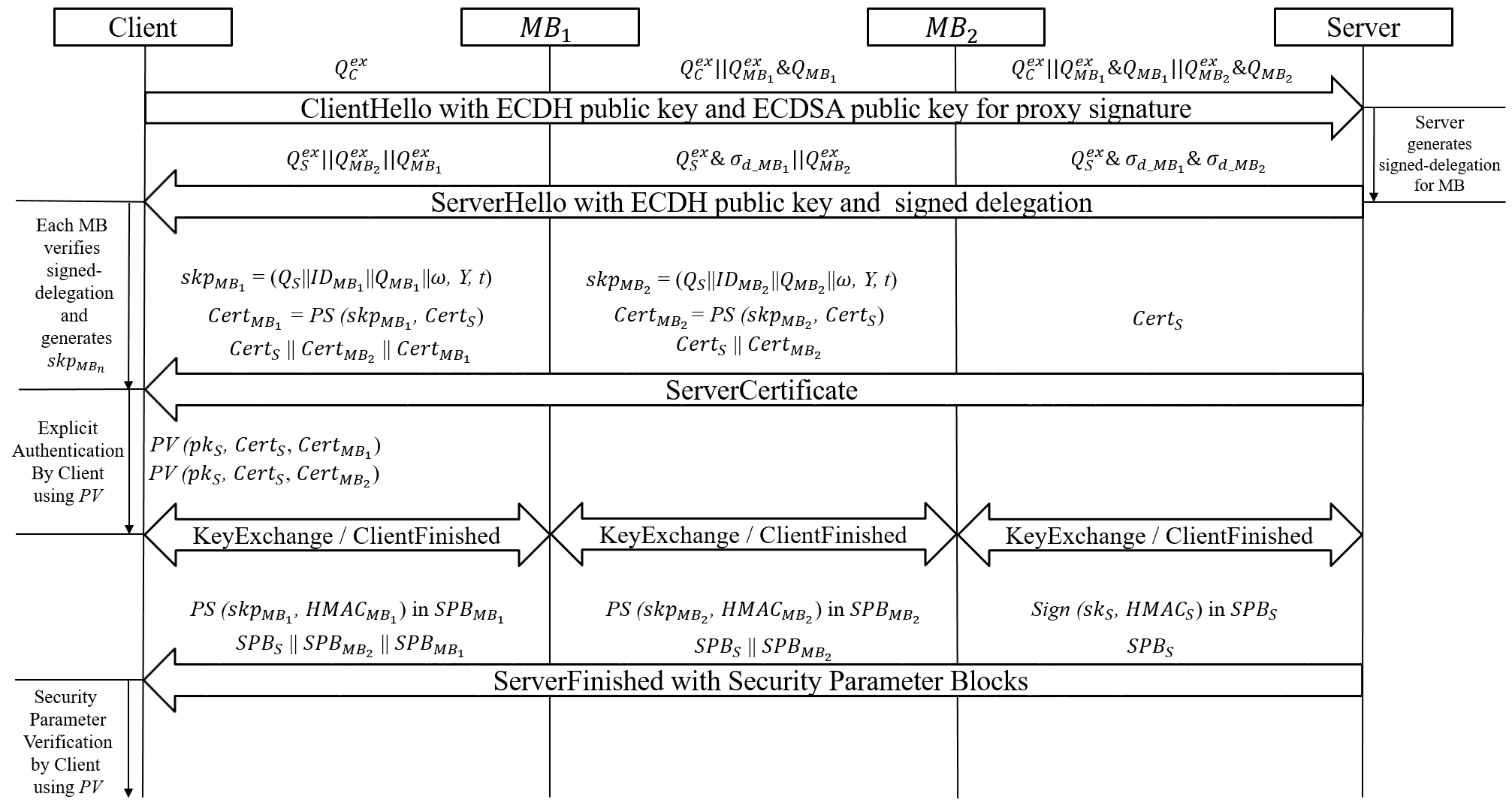}
\caption{Handshake phase of mdTLS} \label{fig1}
\end{figure}

\item Middleboxes attach their own ECDH public key $Q^{ex}_{MB_{i}}$ to the \textsf{ServerHello} message. Then, middleboxes check whether signed delegations from the server are valid. If validation succeeds, middleboxes generate their proxy signing key $skp_{MB_{i}}$.
\begin{itemize}
    \item $skp_{MB_{i}} \leftarrow (Q_{S}||ID_{MB_{i}}||Q_{MB_{i}}||\omega, x_{Y_d}, t)$
    \begin{itemize}
        \item $c \leftarrow H(m_d)$ ($m_d=0||Q_{S}||ID_{MB_{i}}||Q_{MB_{i}}||\omega)$
        \item $r \leftarrow H(Q_{S}||ID_{MB_{i}}||Q_{MB_{i}}||\omega||c)$
        \item $t \leftarrow r + d_{MB_{i}} \cdot H(Y_d||\omega)$ mod $q$
            \begin{itemize}
                \item $Y_d \leftarrow y_d \cdot G = s_d^{-1} \cdot (c + d_{S} \cdot x_{Y_d}) \cdot G$ \vspace{0.1cm}
            \end{itemize}
    \end{itemize}
\end{itemize}

\item Due to the \textsf{ServerCertificate} message, the server sends its certificate $Cert_{S}$ to the client and middleboxes. Middleboxes generate their own certificates $Cert_{MB_{i}}$ by proxy signing the received server's certificate. Then, their certificates are sent to the client by appending to the \textsf{ServerCertificate} message.
\begin{itemize}
        \item $PS(skp_{MB_{i}}, Cert_{S})$ returns $Cert_{MB_{i}}$, which can be shown as below:
        \begin{itemize}
            \item $(ID_{MB_{i}}, Q_{MB_{i}}, \omega, (x_{Y_d}, s_d), S^H(t, 0||Cert_{S}||Q_{S}||ID_{MB_{i}}||Q_{MB_{i}}||\omega||x_{Y_d}||s_d||r))$
                \begin{itemize}
                    \item $ (x_{Y_p}, s_p) \leftarrow S^H(t, 0||Cert_{S}||Q_{S}||ID_{MB_{i}}||Q_{MB_{i}}||\omega||x_{Y_d}||s_d||r)$ \vspace{0.1cm}
                \end{itemize}            
        \end{itemize}
\end{itemize}

\item The client, a verifier, verifies certificates to authenticate entities in TLS session. Unlike $Cert_{S}$, the client has to use proxy verification, $PV$, to verify $Cert_{MB_{i}}$, which requires the client to generate proxy public keys $PKP_{MB_{i}}$ corresponding to each middleboxes. With $PKP_{MB_{i}}$, the client verifies $Cert_{MB_{i}}$.

\begin{itemize}
    \item $PV(Q_{S}, Cert_{S}, Cert_{MB_{i}})$
        \begin{itemize}
            \item $Cert_{MB_{i}} \leftarrow (ID_{MB_{i}}, Q_{MB_{i}}, \omega, (x_{Y_d}, s_d), (x_{Y_p}, s_p))$
            \item If $Cert_{S} \notin \omega $ then \textit{return false}; 
            \item Else $PKP_{MB_{i}} \leftarrow r \cdot G + H(s_d^{-1} \cdot (c \cdot G + x_{Y_d} \cdot Q_{S})||\omega) \cdot Q_{MB_{i}}$;
            \begin{itemize}
                \item $c \leftarrow H(0||Q_{S}||ID_{MB_{i}}||Q_{MB_{i}}||\omega)$,  $r \leftarrow H(Q_{S}||ID_{MB_{i}}||Q_{MB_{i}}||\omega||c)$ 
            \end{itemize}
            \item $V^H(PKP_{MB_{i}}, 0||Cert_{S}||Q_{S}||ID_{MB_{i}}||Q_{MB_{i}}||\omega||x_{Y_d}||s_d||r, (x_{Y_p}, s_p))$ \vspace{0.1cm}
        \end{itemize}
\end{itemize}

\item Server sends \textsf{ServerFinished} message with security parameter block ($SPB$). These blocks consist of signatures of $HMAC$. This $HMAC$ generates authentication code from security parameters such as ciphersuite and handshake messages. For middleboxes, they have to proxy sign their blocks with their generated $skp_{MB_{i}}$. For a client, it must verify middleboxes’ signed blocks with its generated proxy public keys $PKP_{MB_{i}}$.
\end{enumerate}

\subsubsection{Phase 2. Record}
\begin{enumerate}
\item 
Modification log is attached to the message and helps to check whether a message is modified. Besides, endpoints can also check whether unauthorized entities modify messages without permission.
\end{enumerate}
\vspace{-0.6cm}
\begin{figure}
\includegraphics[width=\textwidth]{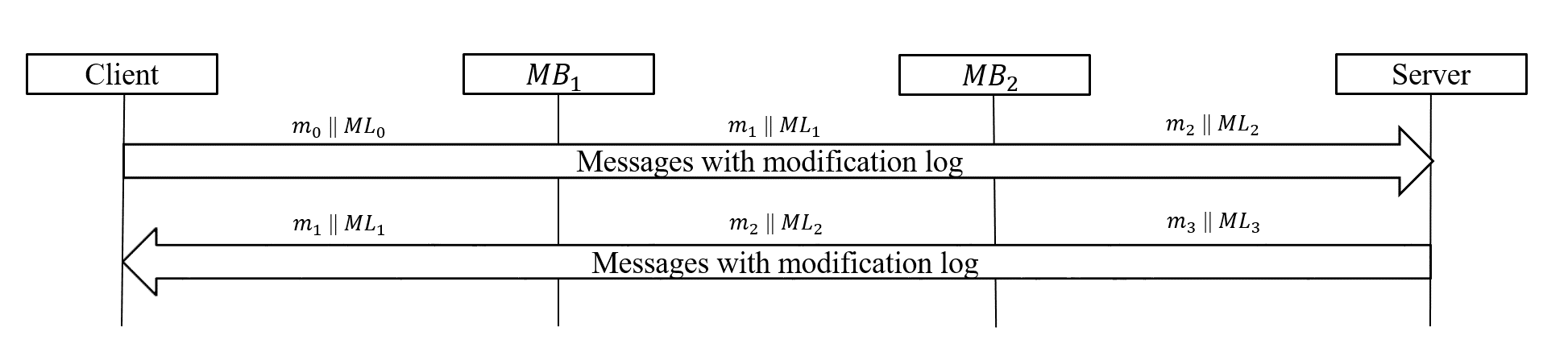}
\caption{Record phase of mdTLS} \label{fig2}
\end{figure}

\vspace{-0.8cm}
\section{Performance analysis for mdTLS}
%In this section, we described the performance of the maTLS and mdTLS protocols based on the number of point multiplications. Unlike maTLS, we did not implement our protocol as a source code. So, we determined performances by analyzing the amount of overall computations in protocols. Since ECDSA is based on ECDLP (Elliptic Curve Discrete Logarithm Problem) and performs point multiplication operations on vast numbers, these point multiplication operations significantly impact performance. Therefore, the performance is analyzed using an algorithm that can measure the number of point multiplication operations. Furthermore, these analyses are based on the TLS 1.2 version with server-only authentication and assume that 3 SCTs are created for each certificate through the Certificate Transparency policy \cite{ref_url10,ref_url9,ref_url11,ref_url16}.
% (기존) 이 섹션에서는 point multiplication 수를 기반으로 maTLS 및 mdTLS 프로토콜의 성능을 설명한다. maTLS와 달리 우리는 프로토콜을 소스 코드로 구현하지 않았다. 그래서 우리는 프로토콜 전체의 연산량을 분석하여 성능을 판단하였다. ECDSA는 ECDLP(타원 곡선 이산 로그 문제)를 기반으로 하고 방대한 수의 point multiplication 연산을 수행하므로 이러한 point multiplication 연산은 성능에 큰 영향을 미친다. 따라서 point multiplication 연산 횟수를 측정할 수 있는 알고리즘을 이용하여 성능을 분석하였다. 이 분석은 server-only authenticcated TLS 1.2 버전을 기반으로 하며 인증서 투명성 정책을 통해 각 인증서에 3개의 SCT를 생성하는 것을 전제로 하였다.
\textcolor{black}{In this section, we analyzed the performance of the mdTLS by conducting a comparative analysis with maTLS, which we consider to be among the best of the existing TLS-extension protocols. Our performance analysis is focused on the number of computations in protocols. Both mdTLS and maTLS rely on ECDSA for the generation of security parameters. ECDSA, being based on the Elliptic Curve Discrete Logarithm Problem (ECDLP), involves a substantial number of point multiplication operations. These operations can significantly influence the performance of both protocols. Therefore, we conducted a performance analysis employing algorithms capable of measuring the number of point multiplication operations. It's important to note that this analysis is based on server-only authenticated TLS version 1.2 and assumes that 3 SCTs are created for each certificate through the Certificate Transparency policy \cite{ref_url10,ref_url9,ref_url11,ref_url16}}.
%the Certificate Transparency policy generates 3 SCTs for each certificate.
% (수정) 이 섹션에서 우리는 mdTLS 프로토콜의 성능이 얼마나 우수한지에 대해 설명한다. 이를 설명하기 위해 우리는 기존 프로토콜 중 가장 우수하다고 판단되는 maTLS와 mdTLS의 성능을 비교하였다. 우리는 설계된 프로토콜 내 연산량을 바탕으로 성능의 우수성을 판단하였다. mdTLS와 maTLS 모두 암호 파라미터를 생성할 때, ECDSA를 바탕으로 한다. ECDSA는 ECDLP(타원 곡선 이산 로그 문제)를 기반으로 하고 방대한 수의 point multiplication 연산을 수행하기 때문에, 두개의 프로토콜 모두 point multiplication 연산이 성능에 큰 영향을 끼질 수 있다. 따라서 우리는 point multiplication 연산 횟수를 측정할 수 있는 알고리즘을 이용하여 성능을 분석하였다. 이 분석은 server-only authenticcated TLS 1.2 버전을 기반으로 하며 인증서 투명성 정책을 통해 각 인증서에 3개의 SCT를 생성하는 것을 전제로 하였다.

\subsection{Preliminaries for performance analysis}
%To provide 128-bit security strength, we matched the sizes of related parameters as below\cite{ref_article13}.

\textcolor{black}{To facilitate performance comparisons between two protocols that offer the same 128-bit security strength, we have set the elements within the protocols, as shown below \cite{ref_article13}.}
%we set the elements within the protocols that can affect security strength to the same value, as shown in the below.}
% (수정) 두개의 프로토콜이 동일한 128비트의 암호 강도를 제공할 때의 성능을 비교하기 위해 우리는 아래와 같이 암호강도에 영향을 끼칠 수 있는 프로토콜 내 요소 등을 동일한 값으로 통일시켰다.
% (수정-2) 동일한 128비트 보안 강도를 제공하는 두 프로토콜 간의 성능 비교를 용이하게 하기 위해 아래와 같이 프로토콜 내에 요소를 설정했습니다.
\begin{itemize}
\item Types of elliptic curve: Secp256r1
\item Secret key size: 256 bits
\item Hash size: 256 bits
\end{itemize}

%Using them, we calculated the number of multiplications in ECDSA and Schnorr signature scheme, as shown in Table 2.
%\begin{table}[h!]
%\caption{Number of multiplications in digital signatures}\label{tab2}
%\centering
%\begin{tabular}{|C{40mm}|C{35mm}|C{35mm}|}
%\hline
 %                          & \begin{tabular}[c]{@{}c@{}}Scalar Multiplications\\ in ECDSA\end{tabular}      & \begin{tabular}[c]{@{}c@{}}Modular Multiplications\\ in Schnorr\end{tabular}       \\
%\hline
%Key Generation             & 384        & 384         \\
%Signature Generation       & 384        & 385           \\
%Signature Verification     & 768        & 448           \\
%\hline
%\end{tabular}
%\end{table}

\subsection{Analyzing the performance between maTLS and mdTLS}
%After setting the parameters, we analyzed the computations of the protocol in three parts by using the double-and-add algorithm. This algorithm requires an average of 1 point doubling and 0.5 point addition per bit. Therefore, we considered 1.5 point multiplications per bit are needed on average.\\
%Among three parts, the overall operations for generating and verifying certificates are shown in Table 3 and 4 below. Table 2 shows operations for security parameter blocks. Summarizing the results in these tables, the overall point multiplications for maTLS and mdTLS are 8,832(N + 1) and 5,376N + 8,832, respectively. N stands for the number of middleboxes. This means that the amount of operations in mdTLS is less than those of maTLS. The operation of mdTLS was reduced by about 39.1\% compared to maTLS. The rationales for deriving this result are shown in Figure 3.
% (기존) 매개변수를 설정한 후 double-and-add 알고리즘을 사용하여 프로토콜 계산을 세 부분으로 나누어 분석했다. 이 알고리즘에서는 비트당 평균 1번의 doubling과 0.5번의 addition을 필요로 한다. 따라서 비트당 평균 1.5 point multiplication의 연산이 발생한다고 간주하였다. \\
%세 부분 중 인증서 생성 및 검증을 위한 전반적인 연산량은 아래 표 3 및 4와 같다. 표 2는 보안 매개변수 블록에 대한 연산량을 보여준다. 이 표들의 결과를 요약하면 maTLS와 mdTLS의 전체 포인트 곱셈 수는 각각 8,832(N + 1) 및 5,376N + 8,832이다. 이 경우 N은 미들박스 수를 나타낸다. 이는 mdTLS의 작업량이 maTLS의 작업량보다 적음을 의미하며, mdTLS의 연산은 maTLS에 비해 약 39.1% 감소했다. 이 결과는 그림 3에 도식화하였다.
\textcolor{black}{To measure the number of point multiplication operations, we employed the double-and-add algorithm, which averages 1 point doubling and 0.5 point additions per bit. Therefore, we considered an average of 1.5 point multiplication operations per bit. Following this, we divided the protocol into two segments and measured the number of point multiplication operations. 
The first segment corresponds to the generation and verification of certificates for utilization in the handshake phase. The number of computations for each protocol in this segment is detailed in Table 3 and 4 below. The second segment is where entities (server, client, middlebox) create and verify security parameters to be exchanged at the handshake phase. The number of computations for each protocol in this segment is detailed in Table 2 below.}
% (수정) point multiplication 연산 횟수를 측정하기 위해 우리는 double-and-add 알고리즘을 사용하였습니다. 이 알고리즘에서는 비트당 평균 1번의 doubling과 0.5번의 addition을 필요로 한다. 따라서 비트당 평균 1.5 point multiplication의 연산이 발생한다고 간주하였다. 이 규칙에 따라 우리는  프로토콜을 크게 두 부분으로 나눈 후, 연산 수를 측정하였다. 첫 번째 부분은 Handshake 단계와 Record phase에서 사용할 인증서를  생성 및 검증하는 부분이다. 해당 부분에서 발생하는 각 프로토콜의 연산량은 아래 표 3 및 4와 같다. 두 번째 부분은 Handshake 단계와 Record phase에서 개체들(서버, 클라이언트, 미들박스)이 메세지를 주고 받을 때, 필요한 보안 매개변수를 생성 및 검증하는 부분이다. 해당 부분에서 발생하는 각 프로토콜의 연산 량은 아래 표 2와 같다. 이 표들의 결과를 모두 종합하면 maTLS와 mdTLS의 전체 포인트 곱셈 수는 각각 8,832(N + 1) 및 5,376N + 8,832이다. 이때 N은 미들박스 수를 나타낸다. 이는 두개의 프로토콜이 동일한 보안강도를 제공한다고 가정할 때, 아래 그림 3과 같이mdTLS에 필요한 연산량이 maTLS보다 약 39.1% 적은 것으로 나타났다.
% (수정-2) point multiplication 연산 횟수를 측정하기 위해 우리는 double-and-add 알고리즘을 사용하였습니다. 이 알고리즘에서는 비트당 평균 1번의 point doubling과 0.5번의 point addition을 필요로 한다. 따라서 비트당 평균 1.5 point multiplication의 연산이 발생한다고 간주하였다. 이 규칙에 따라 우리는  프로토콜을 크게 두 부분으로 나눈 후, 연산 수를 측정하였다. 첫 번째 부분은 Handshake 단계에서 사용할 인증서를  생성 및 검증하는 부분이다. 해당 부분에서 발생하는 각 프로토콜의 연산량은 아래 표 3 및 4와 같다. 두 번째 부분은 개체들(서버, 클라이언트, 미들박스)이 Handshake 단계에서 주고 받을 보안 매개변수를 생성 및 검증하는 부분이다. 해당 부분에서 발생하는 각 프로토콜의 연산 량은 아래 표 2와 같다. 이 표들의 결과를 모두 종합하면 maTLS와 mdTLS의 전체 포인트 곱셈 수는 각각 8,832(N + 1) 및 5,376N + 8,832이다. 이때 N은 미들박스 수를 나타낸다. 이는 두개의 프로토콜이 동일한 보안강도를 제공한다고 가정할 때, 아래 그림 3과 같이mdTLS에 필요한 연산량이 maTLS보다 약 39.1% 적은 것으로 나타났다.
\vspace{-0.5cm}
\begin{table}[h!]
\caption{Computational analysis for security parameter blocks}\label{tab4}
\centering
\begin{tabular}{|L{85mm}|C{15mm}||C{15mm}|}
\hline
\multicolumn{1}{|c|}{Descriptions} & \begin{tabular}[c]{@{}c@{}}maTLS\end{tabular} & \begin{tabular}[c]{@{}c@{}}mdTLS\end{tabular}       \\
\hline
{\scriptsize Server generates security parameter blocks.} & 384 & 384   \\
{\scriptsize Middlebox generates security parameter blocks.} & 384N & 384N   \\
{\scriptsize Client verifies blocks from the server.} & 768 & 768   \\
{\scriptsize Client verifies blocks from the middleboxes.} & 768N & 768N   \\
\hline
\end{tabular}
\end{table}

\begin{table}[h!]
\caption{{Computational analysis for generating certificates}}\label{tab2}
\centering
\begin{tabular}{|L{85mm}|C{15mm}||C{15mm}|}
\hline
\multicolumn{1}{|c|}{Descriptions} & \begin{tabular}[c]{@{}c@{}}maTLS\end{tabular} & \begin{tabular}[c]{@{}c@{}}mdTLS\end{tabular}       \\

\hline
\textbf{- Server side} &  & \\
\;{\scriptsize Server generates keys and signature for CSR to CA.} & 768 & 768       \\
\;{\scriptsize CA verifies CSR signature.} & 768 & 768       \\
\;{\scriptsize CT log servers generate keys and signatures for 3 SCTs.} & 2,304 & 2,304     \\
\;{\scriptsize CA generates keys and signs for server's certificate.} & 768 & 768     \\
\hline
\textbf{- Middlebox side for maTLS} &  & \\
\;{\scriptsize Middleboxes generate keys and signature for CSR to CA.} & 768N & -  \\
\;{\scriptsize CA verifies CSR signature.} & 768N & -  \\
\;{\scriptsize MT log servers generate keys and signatures for 3 SCTs.} & 2,304N & - \\
\;{\scriptsize CA generates keys and signs for middleboxes' certificate.} & 768N & -  \\
\hline
\textbf{- Middlebox side for mdTLS} &  & \\
\;{\scriptsize Each middlebox generates its keys.} & - & 384N  \\
\;{\scriptsize Server generates signed delegations to assign proxy signers.} & - & 384N  \\
\;{\scriptsize Middlebox verifies signed delegation and generate proxy signing key.} & - & 768N \\
\;{\scriptsize Middleboxes generate certificates with proxy signing key.} & - & 384N  \\
\hline
\end{tabular}
\end{table}

\vspace{-0.8cm}

\begin{table}[h!]
\caption{{Computational analysis for certificates verification}}\label{tab3}
\centering
\begin{tabular}{|L{85mm}|C{15mm}||C{15mm}|}
\hline
\multicolumn{1}{|c|}{Descriptions} & \begin{tabular}[c]{@{}c@{}}maTLS\end{tabular} & \begin{tabular}[c]{@{}c@{}}mdTLS\end{tabular}       \\

\hline
\;{\scriptsize Client verifies the signature and 3 SCTs in the server's certificate.} & 3,072 & 3,072       \\
\;{\scriptsize Client verifies the middleboxes' certificates.} & 3,072N & 2,304N       \\
\hline
\end{tabular}
\end{table}

\vspace{1cm}

\begin{figure}[h!]
     \begin{subfigure}[b]{0.5\textwidth}
         \centering
         \includegraphics[width=\textwidth, height=3.3cm]{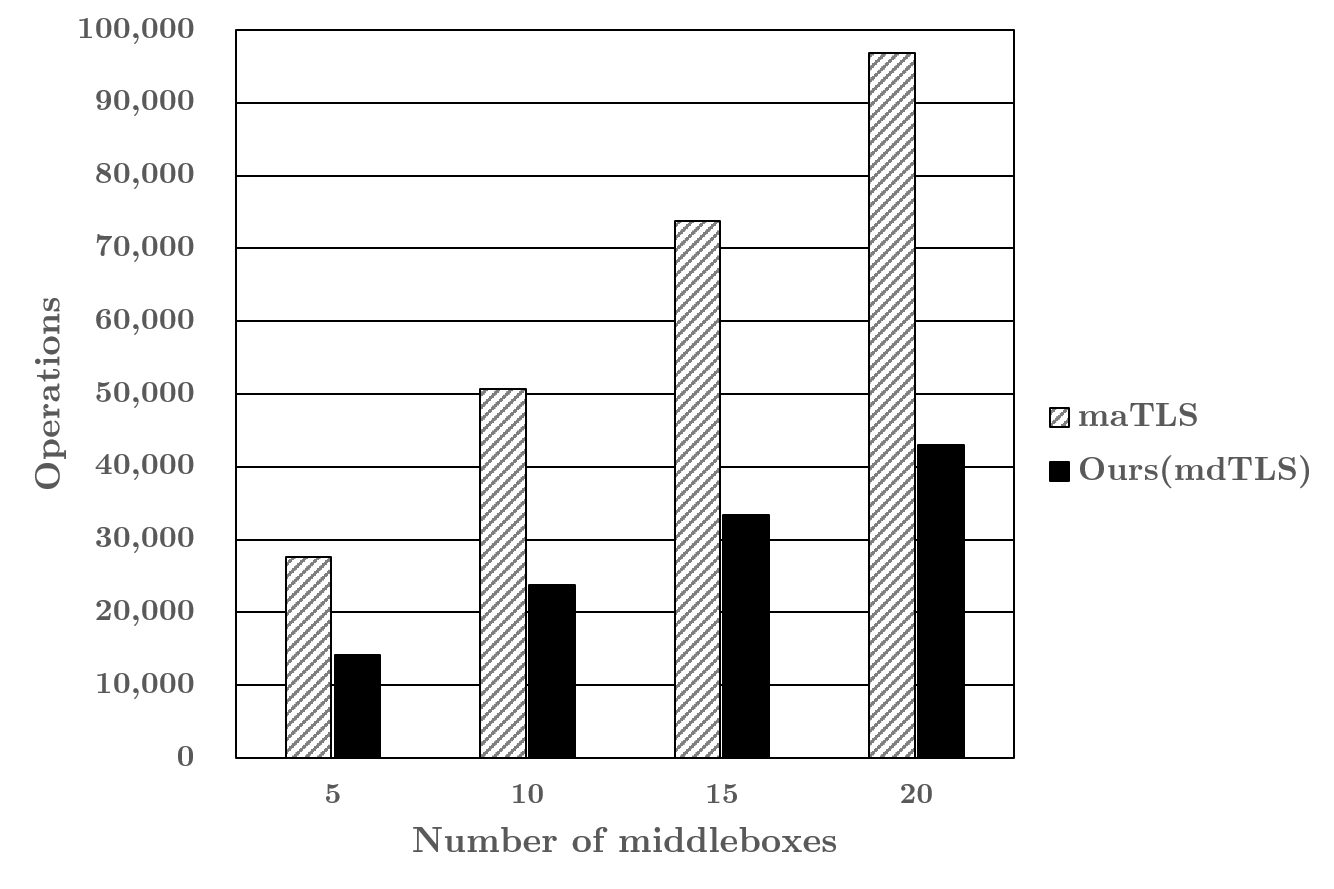}
         \caption{Certificate generation}
         \label{Certificate Generation}
     \end{subfigure}
     \hfill
     \begin{subfigure}[b]{0.5\textwidth}
         \centering
         \includegraphics[width=\textwidth, height=3.3cm]{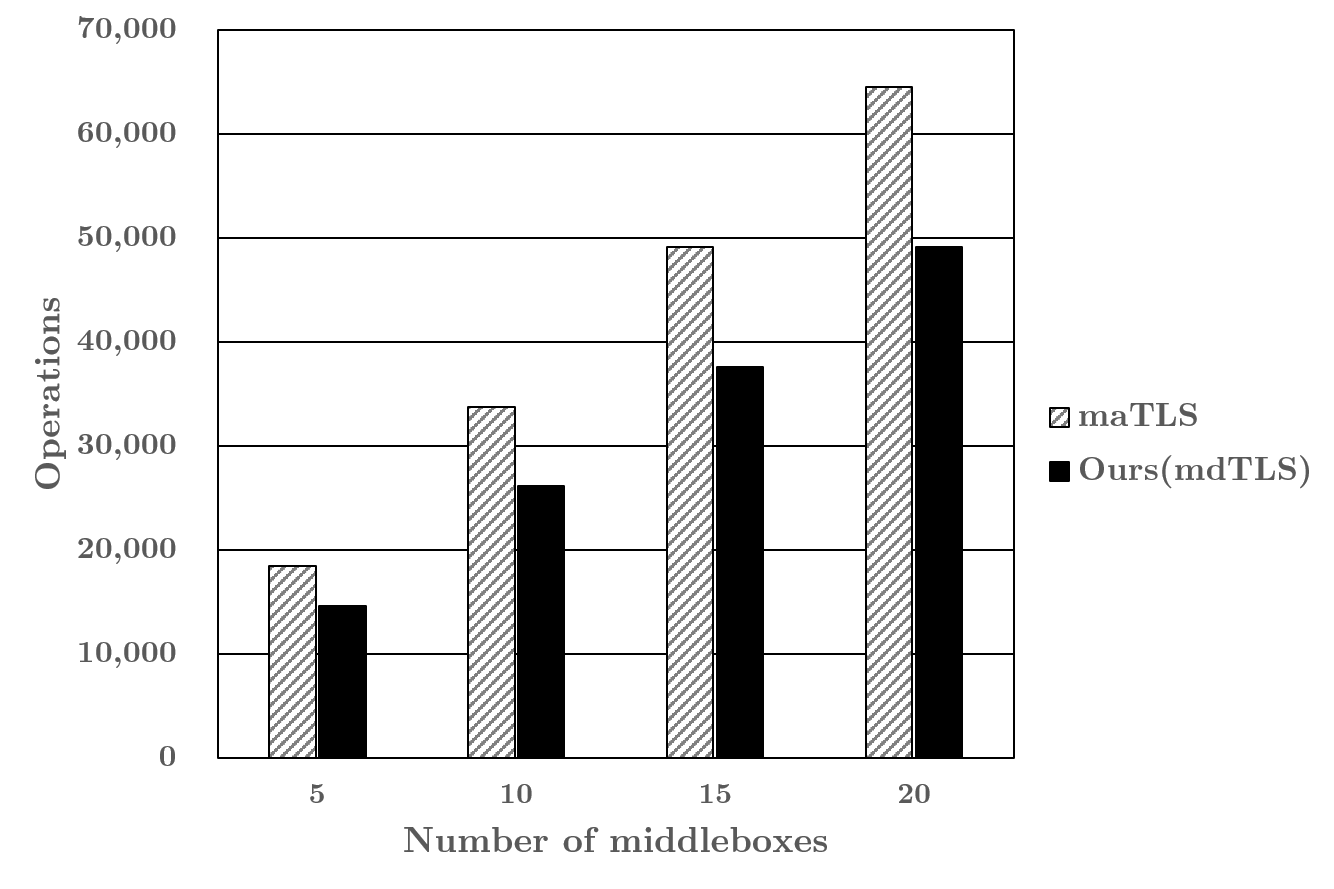}
         \caption{Certificate verification}
         \label{Certificate Verification}
     \end{subfigure}
     \hfill
     \begin{subfigure}[b]{0.5\textwidth}
         \centering
         \includegraphics[width=\textwidth, height=3.3cm]{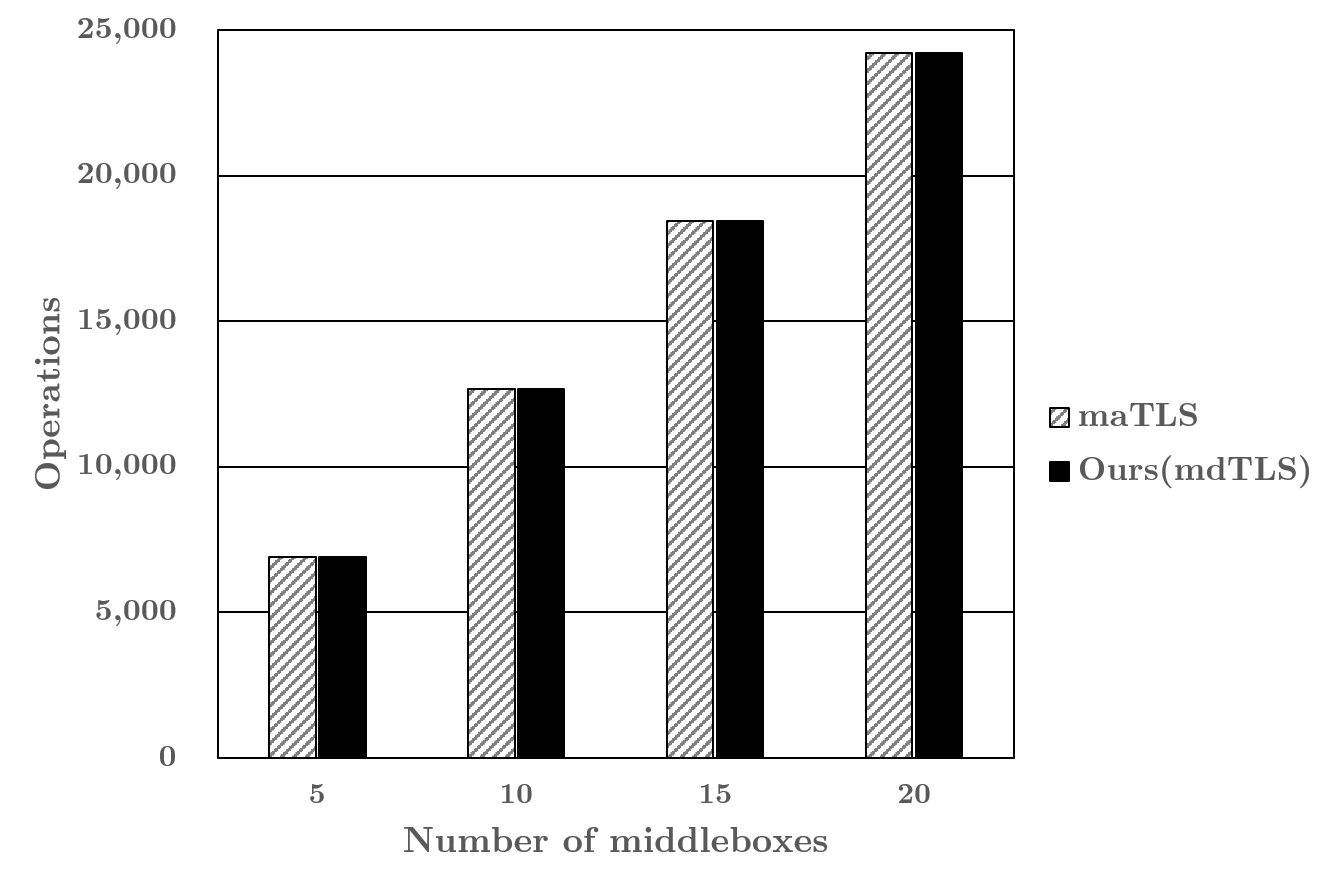}
         \caption{Security parameter blocks}
         \label{Security Parameter Blocks}
     \end{subfigure}
     \hfill
     \begin{subfigure}[b]{0.5\textwidth}
         \centering
         \includegraphics[width=\textwidth, height=3.3cm]{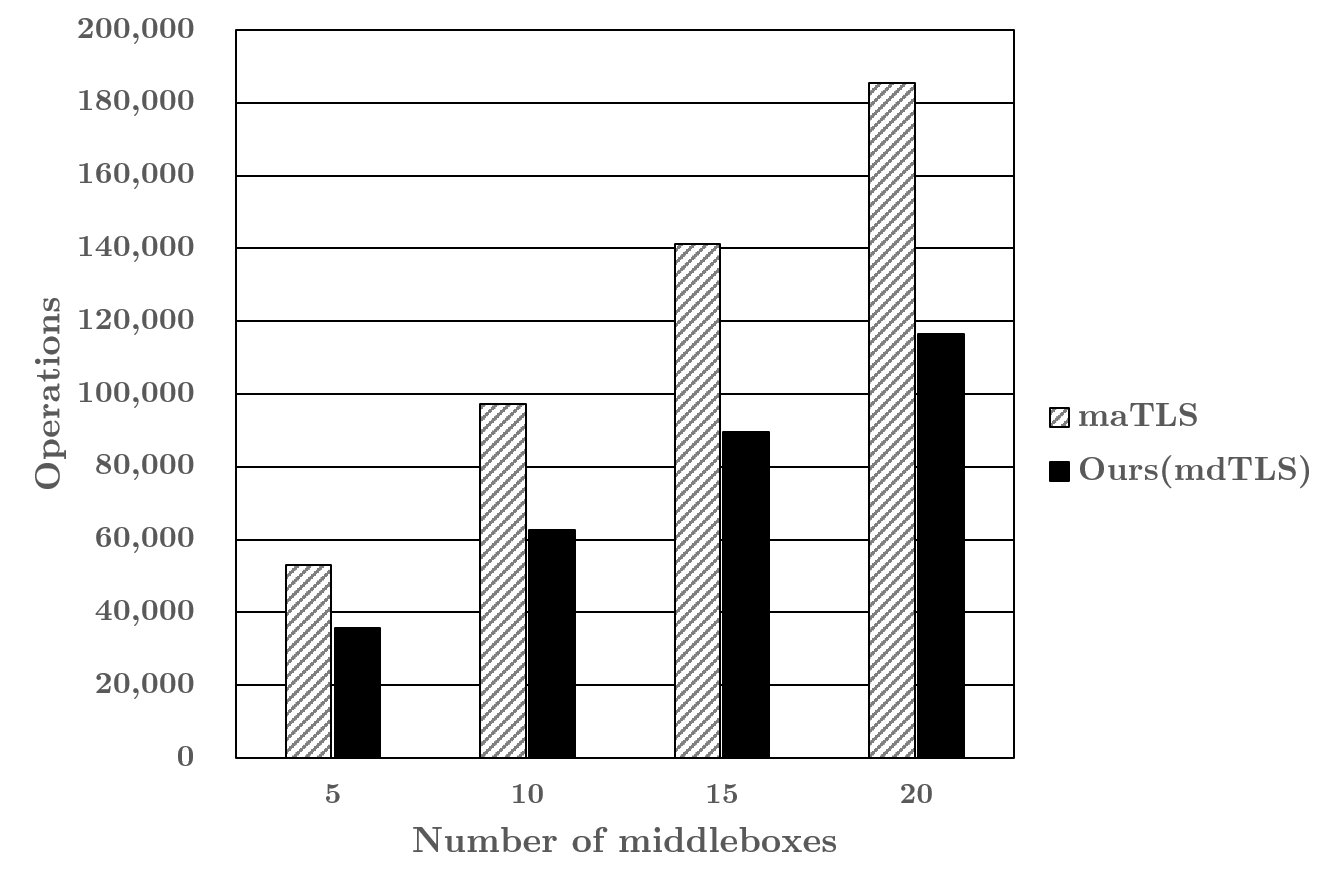}
         \caption{Overall performances}
         \label{Overall Performances}
     \end{subfigure}
        \caption{Performance of protocols when using ECDSA}
        \label{fig: graphs}
\end{figure}
\vspace{-0.8cm}

\newpage

\section{Security analysis for mdTLS}

In this section, we conducted a security analysis of mdTLS using an approach similar to the one employed for maTLS \cite{ref_proc8}, involving formal specification and verification through the Tamarin \cite{ref_url7}. Tamarin is an automated formal verification tool based on multiset rewriting rules in the theory of equations. It has been continuously updated to maintain its effectiveness. Using this tool, maTLS formally verified six security lemmas: \textit{server authentication}, \textit{middlebox authentication}, \textit{data authentication}, \textit{path integrity}, \textit{path secrecy}, and \textit{modification accountability}. In the case of mdTLS, we successfully verified not only the same lemmas as previously done in maTLS but also three novel lemmas related to the proxy signature scheme following the same approach and tools: \textit{verifiability}, \textit{strong-unforgeability}, and \textit{strong-identifiability}. However, in this paper, we only described three novel lemmas related to the proxy signature scheme, taking into consideration the maximum page limit imposed by the conference guidelines. The rest can be found on our GitHub \cite{ref_url12}.

\subsection{Experimental setup}
\textcolor{black}{To analyze the security of the mdTLS, we established an experimental environment, as illustrated below. Our goal was to confirm that the formal model of mdTLS aligns with the security lemmas within our testing environment.}
%mdTLS 프로토콜의 보안성을 평가하기 위해 아래와 같이 실험 환경을 설정했다. 우리는 maTLS와 mdTLS의 formal specification이 해당 환경에서 보안 lemma를 충족하는지 확인하고자 하였다.
\begin{itemize}
\item Amazon Elastic Compute Cloud (Amazon EC2) c5a.24xlarge instance
\item 96 vCPUs, 192 GiB memories
\item Ubuntu 22.04.2 LTS
\end{itemize}

\subsection{Formal specification}

\textcolor{black}{We have formalized the mdTLS, specifying the detailed operations conducted by each entity during the handshake and record phases in the form of rules. For cryptographic primitives like hash, signature, and PRF (Pseudo-Random Function) \cite{ref_article14}, we used the built-in functions provided by Tamarin. Details of all rules can be found in the spthy file uploaded to our github \cite{ref_url12}. The script below illustrates an example of the detailed operations concerning \textsf{ServerHello} messages. In the handshake phase, when the server receives a \textsf{ClientHello} message from the client, it responds by sending a \textsf{ServerHello} message to initiate mutual authentication. In this process, mdTLS sends a \textsf{ServerKeyExchange} message, a signed delegation, a Diffie-Hellman public key, and a \textsf{ServerCertificate} message. The delegation in this context consists of the server's public key, the middlebox's public key and identification information, and a warrant providing an explanation of the delegation.}
% (수정) 우리는 mdTLS 프로토콜을 정형명세하였다. 프로토콜 내 handshake , record phase 내 각 개체들이 수행하는 세부 작업들은 rule 형태로 정형하게 명세되며 해시, 서명, PRF(Pseudo-Random Function)와 같은 암호 프리미티브의 경우, Tamarin에서 제공하는 내장 기능을 활용하였다.모든 rule의 세부정보는 GitHub\cite{ref_url12}에 업로드된 spthy 파일에서 확인할 수 있으며, 아래 그림은 ServerHello 메시지 관련 세부 작업을 정형하게 명세한 예이다. Handshake phase에서 서버가 클라이언트로부터 ClientHello 메시지를 받는 경우, 상호 인증을 수행하기 위해 ServerHello 메시지를 전송한다. 이때 mdTLS는 특정 미들박스를 프록시 서명자로 지정하기 위해 ServerKeyExchange 메시지, signed delegation, accountability key를 위한 DH public key, 그리고 ServerCertificate 메시지를 보낸다.이때 위임은 서버의 공개키, 미들박스의 공개키와 그 식별정보, 위임에 대한 설명이 포함된 warrant로 구성된다.
% (수정-2) 우리는 mdTLS를 정형명세하였다. 프로토콜 내 handshake , record phase 내 각 개체들이 수행하는 세부 작업들은 rule 형태로 정형하게 명세되며 해시, 서명, PRF(Pseudo-Random Function) \cite{ref_article14}와 같은 암호 프리미티브의 경우, Tamarin에서 제공하는 내장 기능을 활용하였다.모든 rule의 세부정보는 GitHub\cite{ref_url12}에 업로드된 spthy 파일에서 확인할 수 있으며, 아래 그림은 ServerHello 메시지 관련 세부 작업을 정형하게 명세한 예이다. Handshake phase에서 서버가 클라이언트로부터 ClientHello 메시지를 받는 경우, 상호 인증을 수행하기 위해 ServerHello 메시지를 전송한다. 이때 mdTLS는 특정 미들박스를 프록시 서명자로 지정하기 위해 ServerKeyExchange 메시지, signed delegation, Diffie-Hellman public key, 그리고 ServerCertificate 메시지를 보낸다.이때 위임은 서버의 공개키, 미들박스의 공개키와 식별정보, 위임에 대한 설명이 포함된 warrant로 구성된다.

\newpage

\begin{mdframed}
\begin{scriptsize}
\begin{verbatim} 
rule Server_Hello:
  let
        server_hello_msg 
            = < 'server_hello', ~ns, server_chosen_details >
                        ...
        server_key_exchange = s_dhe_pub
        server_key_exchange_signed 
            = < server_key_exchange, sign(h(server_key_exchange)
                , ltk) >

        server_cert = < $S, pk(ltk) >
        
        warrant = ~warrant_fresh
        proxy_delegation = < pk(ltk), $M, mb_pubkey, warrant >
        proxy_delegation_signed = sign(h(proxy_delegation), ltk)

        Y_d = calcY_d(~y, 'G_skp')
        y_d_x = pointx(Y_d)
        c = h(proxy_delegation)
        s_d = multp( plus(multp(ltk, y_d_x), c), inv(~y) )
        
        proxy_delegation_signed_pair 
         = < proxy_delegation, proxy_delegation_signed, <y_d_x, s_d> >
  in
    [   In( <mb_client_hello_msg, c_mb_extension> )
        , !PrivateKey('server', $S, ltk)    ]
  --[
        ServerSendDelegation(ltk_pub, mb_pubkey, warrant
                             , proxy_delegation)
    ]->
    [   Out( <server_hello_msg, server_key_exchange_signed
              , proxy_delegation_signed_pair, s_extension
                , server_cert> ) ...    ]
\end{verbatim}
\end{scriptsize}
\end{mdframed}

\subsection{Formal verification}
%We defined nine security lemmas and one source lemma for security verification. Six security lemmas are from Lee et al. \cite{ref_proc8}, and three other security lemmas are newly added to prove the security property of proxy signature, which are \textit{Verifiability}, \textit{Strong-Unforgeability}, and \textit{Strong-Identifiability}. They are defined as first-order logic based formulas called lemma. In this paper, we will discuss the newly added three security properties of the proxy signature scheme.
%보안 검증을 위해 9개의 보안 lemma와 1개의 source lemma를 정의하였다. 6개의 보안 lemma는 Lee et al.이 정의하였고, 프록시 서명의 보안 lemma를 증명하기 위해 Verifiability, Strong-Unforgeability 및 Strong-Identifiability라는 세 가지 다른 보안 lemma를 새로 추가하였다. 이들은 lemma라고 불리는 first-order logic 기반 공식으로 정의된다. 본 논문에서는 새로 추가된 프록시 서명의 세 가지 보안 lemma에 대해 논의한다.

We defined nine security lemmas and one source lemma for security verification. Among them, security lemmas consist of six security lemmas of maTLS and three security lemmas related to the proxy signature scheme. As previously noted, we described three security lemmas associated with proxy signatures. Prior to describing them, we described an additional description of a source lemma designed to assist Tamarin in accurately verifying the formal specifications.
%보안 검증을 위해 9개의 보안 보조 정리와 1개의 소스 보조 정리를 정의했습니다. 9개의 보안 정리는 maTLS의 보안 정리 6개와 프록시 서명 방식과 관련된 보안 정리 3개로 구성됩니다. 앞서 언급했듯이 본 논문에서는 보안 보조정리 중 프록시 서명과 관련된 세 가지 보안 보조정리를 논의한다. 또한 그외 보안 보조정리를 검증하기 전에,  tamarin 도구가 정형모델을 정확히 검증하는데 도움을 줄 수 있는 source lemma를 추가적으로 설명하였다. 앞선 보조정리들은 1차 논리 기반 수식으로 정의합니다.

%\newpage

\paragraph{Source lemma.}
\textcolor{black}{A source lemma is a concept used for formally verifying the security lemmas that a security protocol must adhere to during its execution. When conducting formal verification of an overall protocol, Tamarin adopts a strategy of deconstructing the protocol into smaller, more manageable components for analysis. The verification outcomes for these individual subsets are then used as supporting evidence to confirm that the entire protocol operates correctly and meets its prescribed security lemmas. However, during the verification process of these subsets, if Tamarin encounters difficulties in distinguishing between variables as nonce values or ciphertexts, it may face challenges in completing the verification. This is commonly referred to as a "partial deconstruction". To address such issues, it becomes necessary to establish a source lemma that precisely specifies the origin of these variables. From this source lemma, a refined source is generated, comprising a new set of sources. All security lemmas are subsequently verified using these refined sources, underscoring the importance of validating the source lemma to ensure the accurate computation of these refined sources \cite{ref_proc11,ref_url7}. When we initially omitted the definition of source lemmas, the formally specified mdTLS model yielded 120 partial deconstructions. Consequently, we defined source lemmas to enable Tamarin to discern the origins of these problematic variables. Upon closer analysis, it was determined that the issue of partial deconstruction occurred in 4 distinct segments, one of which pertained to the scenario where a middlebox received an encrypted request message sent by the client. To resolve this particular issue, we formulated a source lemma indicating that the encrypted message \texttt{enc} received by the middlebox had been transmitted from the client through the \texttt{OutClientRequest()} action, as shown below. By employing this approach, we could generate refined sources in a state of "deconstructions completed". This strategic use of source lemmas proved instrumental in addressing the partial deconstruction challenge and facilitating the successful verification process within the mdTLS model.}
% (수정) 소스 렘마는 보안 프로토콜 또는 시스템의 특정 rule이 실행되는 중 만족해야 하는 보안 속성 또는 조건을 공식적으로 검증하는데 사용되는 중요한 개념입니다. Tamarin은 정형하게 명세된 전체 프로토콜에 대한 정형검증을 수행하기 위해, 프로토콜을 여러 작은 단위로 나누어 검증합니다.이러한 하위 집합에 대한 검증 결과는 전체 시스템이 보안 목표를 만족한 채, 올바르게 동작하는 근거로 활용됩니다. 그러나, 이런 하위 집합을 검증하는 과정에서 집합 내 일부 변수가 nonce 값인지 암호문인지 여부를 식별하지 못하는 경우 Tamarin은 검증을 완료하지 못한다. 이를 partial deconstruction 이슈라 한다. 이러한 partial deconstruction left 이슈를 해결하기 위해서는 변수들의 출처를 정확히 명시하기 위한 목적의 source lemma를 작성해야 하며, 이 source lemma를 통해 새로운 source들의 집합으로 구성된 refined source를 만들어내야 한다. 모든 보안 lemma는 이러한 refined sources를 사용하여 검증되므로 이러한 refined sources가 올바르게 계산되었는지 확인하려면 source lemma의 검증이 필수적이다. 우리가 처음에 source lemma를 정의하지 않았을 때, 정형하게 명세된 mdTLS 모델에서 120개의 partial deconstructions가 발생하였다. 따라서 우리는 Tamarin이 이 문제되는 변수의 출처를 알 수 있도록 source lemma를 작성했다. 분석 결과, Partial deconstruction은 총 4 부분에서 발생했으며, 그 중 하나는 클라이언트가 보낸 암호화된 요청 메시지를 미들박스가 수신했을 때 발생했다. 따라서 우리는 미들박스가 수신한 암호화된 메시지 enc가 클라이언트로부터 OutClientRequest()에 의해 전송되었다는 source lemma를 작성했다. 이러한 방법으로 deconstructions completed 상태의 refined sources를 만들 수 있었다.

\begin{mdframed}
\begin{scriptsize}
\begin{verbatim} 
All enc msg #i.
      InMbClientRequest( enc, msg ) @ i
      ==> (Ex #j. KU(msg) @ j & j < i)
           | (Ex #j. OutClientRequest( enc ) @ j & j<i)
\end{verbatim}
\end{scriptsize}
\end{mdframed}

\paragraph{Security lemma.}
% (기존)After eliminating partial deconstructions, we verified that our protocol meets nine security goals using security lemmas. Security goals mentioned in section 3.2 are converted into each lemma which is a mathematical formula. \textcolor{red}{We prove that the lemma is valid in formal specifications.}
% (기존)Partial deconstructino을 제거한 후 우리는 프로토콜이 security lemma를 사용하여 9가지 보안 목표를 충족하는지 확인했다. 3.2절에서 언급한 보안 목표는 mathematical formula 형식인 lemma로 변환된다.
\textcolor{black}{After resolving the partial deconstruction issue, we verified that our protocol meets the nine security lemmas outlined in Section 3.2. In this section, we define three of the nine security goals related to proxy signature scheme. We also defined detailed information about the formulas that convert informal definitions into mathematical formulas called lemmas.}
% (수정) Partial deconstruction들을 제거한 후 우리는 우리 프로토콜이 앞서 3.2절에서 언급한 9가지 보안 목표를 충족하는지에 대해 Tamarin을 활용하여 검증하였다. 이 절에서는 앞서 언급한 9가지 보안 목표 중 프록시 서명과 관련된 3가지에 대해 상세히 정의하고, 비공식적인 정의를 수학 공식으로 변환하는 수식에 대한 정보를 자세히 설명합니다..
\paragraph{- Verifiability:}The client must verify whether the middlebox's certificate, the proxy signature, was created with the consent of the server. To verify this lemma, we have to check whether the middlebox generated its certificate based on delegation and warrant sent by the server through the \textsf{ServerHello} message, as specified in \textit{rule Server\_Hello}.
%클라이언트는 미들박스의 인증서인 대리 서명값이 서버의 동의를 받아 생성되었는지 확인해야 한다. 이 lemma를 확인하려면 rule Server_Hello에 명세된 대로 ServerHello 메시지를 통해 서버가 보낸 delegation과 warrant를 기반으로 미들박스가 인증서를 생성했는지 확인해야 한다.

\begin{mdframed}
\begin{scriptsize}
\begin{verbatim} 
All warrant mbLtk mbCert #tc.
    ClientReceivedProxySign(warrant, pk(mbLtk), mbCert) @tc  
    ==> Ex delegation gy #tmb.
        MbGenerateProxySign(delegation, mbLtk, gy, warrant, mbCert) 
        @tmb & KU(gy) @tmb & not(Ex #tmb. KU(mbLtk) @tmb)   
        ==> Ex sPub #ts.
            ServerSendDelegation(sPub, pk(mbLtk), warrant, delegation) 
            @ts & (#ts < #tmb) & KU(sPub) @ts
\end{verbatim}
\end{scriptsize}
\end{mdframed}

%\newpage

\paragraph{- Strong-unforgeability:}
The proxy signer’s secret key, which is used to generate the proxy signature, must not be revealed. Otherwise, the proxy signature can be forged by an adversary.
%대리 서명을 생성하는 데 사용되는 대리 서명자의 비밀 키가 공개되면 안되며, 공개될 경우 대리 서명값이 공격자에 의해 위조될 수 있다.
\newpage
\begin{mdframed}
\begin{scriptsize}
\begin{verbatim} 
All warrant mbLtk mbCert #tc.
    ClientReceivedProxySign(warrant, pk(mbLtk), mbCert) @tc
    ==> All delegation gy sPub #tmb.
        (MbGenerateProxySign(delegation, mbLtk, gy, warrant, mbCert) @tmb
         & KU(gy)@tmb & not(Ex #tmb.KU(mbLtk) @tmb)) 
        & (MbReceiveProxyDelegation(sPub, pk(mbLtk), delegation) @tmb)
        ==> All #ts. 
            ServerSendDelegation(sPub, pk(mbLtk), warrant, delegation)@ts & KU(sPub)@ts
            ==> Ex #tmbclient. MbSendPublicKey(pk(mbLtk)) @tmbclient 
                & KU(pk(mbLtk)) @tmbclient
\end{verbatim}
\end{scriptsize}
\end{mdframed}

\paragraph{- Strong-identifiability:}
The identification of a proxy signer can be proved by its public key. The public key of the middlebox included in the proxy signature sent to the client must be the same as the public key of the middlebox sent to the server for proxy delegation.
%대리서명자의 신원은 대리서명자의 공개 키로 증명할 수 있다. 클라이언트에게 전송된 대리 서명에 포함된 미들박스의 공개키는 대리서명 위임을 위해 서버에 전송된 미들박스의 공개키와 동일해야 한다.

\begin{mdframed}
\begin{scriptsize}
\begin{verbatim} 
All warrant mbPub mbCert #tc.
    ClientReceivedProxySign(warrant, mbPub, mbCert)@tc
    ==> All delegation mbLtk gy sPub #tmb.
        (MbGenerateProxySign(delegation, mbLtk, gy, warrant, mbCert)
          @tmb & KU(gy)@tmb & not(Ex #tmb. KU(mbLtk) @tmb))
        & (MbReceiveProxyDelegation(sPub, pk(mbLtk), delegation) @tmb)
        ==> All #ts. 
            ServerSendDelegation(sPub, pk(mbLtk), warrant, delegation)
            @ts & KU(sPub)@ts
            ==> Ex #tmbclient. MbSendPublicKey(pk(mbLtk)) @tmbclient 
                & KU(pk(mbLtk)) @tmbclient & (mbPub = pk(mbLtk))
\end{verbatim}
\end{scriptsize}
\end{mdframed}

%\subsection{Formal Verification}
%% (기존) \subsection{Results of verification}
%% (기존)We verified nine security lemmas and one source lemma using Tamarin. Figure 4 shows that the three security lemmas previously introduced in Section 5.3 and the lemmas proven in maTLS are also satisfied in our mdTLS. Figure 5 shows that, among security lemmas stated in Figure 4, the lemma Verifiability is always true in mdTLS
%(기존)그림 4는 이전에 섹션 5.2에서 소개된 세 가지 보안 lemma와 maTLS에서 입증된 lemma가 우리의 mdTLS 프로토콜에서도 충족된다는 것을 보여준다. 그림 5는 그림 4에 명시된 보안 lemma 중 Verifiability lemma가 mdTLS 프로토콜에서 항상 참임을 보여줍니다.

\paragraph{Results of verification} The overall result of formal verification is shown in Figure 4. Figure 4 illustrates that our mdTLS protocol not only satisfies the three security lemmas introduced above but also aligns with the lemmas validated for maTLS. Furthermore, Figure 5 shows mathematical proofs (verification process) demonstrating the consistent validity of the \textit{verifiability} lemma within our mdTLS protocol among the security lemmas outlined in Figure 4.
% (수정) 섹션 5.2에서 언급한 공식적으로 지정된 mdTLS 프로토콜이 앞서 언급한 보안 목표를 만족하는지 검증한 결과는 아래 그림 4에 나와 있습니다. 그림 4는 앞서 5.2절에서 소개한 세 가지 보안 정리와 maTLS에 대해 증명된 정리가 mdTLS 프로토콜에서도 만족됨을 보여줍니다. 그림 5는 그림 4에서 설명한 보안 정리 중 검증 가능성 정리가 mdTLS 프로토콜에서 항상 참임을 증명하기 위한 수학적 증명(증명 과정)을 보여줍니다.
% (수정-2) 위에서 언급한 공식적으로 지정된 mdTLS 프로토콜이 앞서 언급한 보안 목표를 만족하는지 검증한 결과는 아래 그림 4에 나와 있습니다. 그림 4는 앞서 소개한 세 가지 보안 정리와 maTLS에 대해 증명된 정리가 mdTLS 프로토콜에서도 만족됨을 보여줍니다. 그림 5는 그림 4에서 설명한 보안 정리 중 검증 가능성 정리가 mdTLS 프로토콜에서 항상 참임을 증명하기 위한 수학적 증명(증명 과정)을 보여줍니다.

\vspace{-0.5cm}
\begin{figure}
\centerline{\includegraphics[width=\textwidth]{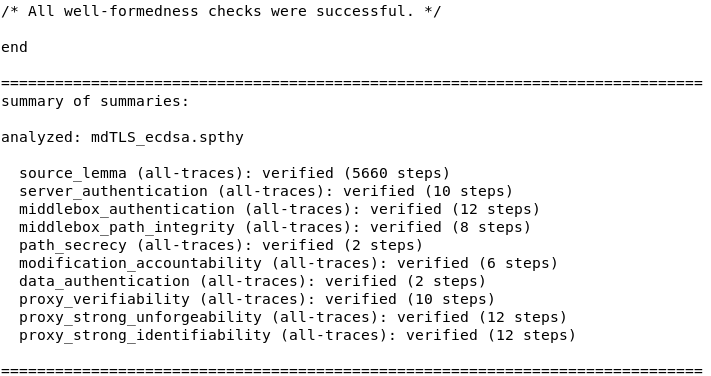}}
\caption{Overview of formal verification results} \label{fig1}
\end{figure}

%\vspace{-1.0cm}

%width=7cm, height=7cm
\begin{figure}
\centerline{\includegraphics[width=9cm, height=8cm]{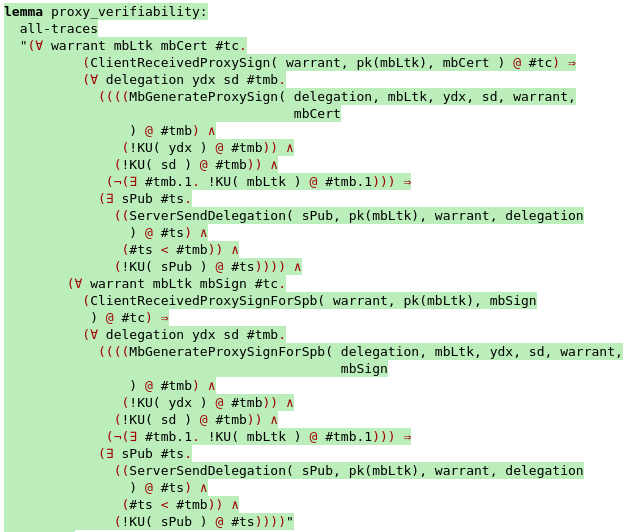}}
\caption{Proof of verifiability lemmas in Tamarin} \label{fig1}
\end{figure}

\vspace{-0.7cm}

\section{Discussion}
\textcolor{black}{We proposed an ECDSA-based cryptographic protocol. However, during the research, we found new insights for improvement. The insight is to use the Schnorr algorithm instead of ECDSA for the algorithm that generates the digital signature.} Boldyreva et al.'s research \cite{ref_article5} used Schnorr signature, and they shows better outcomes in terms of both performance and security than ECDSA. 
\vspace{-0.2cm}
\begin{itemize}
    \item Performance: Schnorr does not have modular inverse calculations that significantly affect performance.
    \item Security: Since Schnorr is strongly unforgeable under chosen message attack (SUF-CMA), Schnorr is provably secure in the random oracle model \cite{ref_article6}.
\end{itemize}

So we compared the performance of the maTLS and mdTLS protocols assumed that both protocols use the Schnorr signature. To measure the performance of Schnorr, the number of modular multiplication operations was calculated using the square-and-multiply algorithm. This algorithm requires 1.5 modular multiplications per bit on average. Besides, as mentioned in Schnorr's paper \cite{ref_proc9}, we calculated the modular multiplications of the Schnorr verification equation by multiplying by 1.75 per bit. When the security level is set to 128-bit, the related parameters' sizes can be shown below \cite{ref_article13}.
\begin{itemize}
\item Public key size: 3,072 bits
\item Secret key size: 256 bits
\item Hash size: 256 bits
\end{itemize}

Table 5 shows the number of modular multiplications at each stage. Here, N represents the number of middleboxes. The mdTLS reduces the number of modular multiplications by 51.8\% compared to maTLS, demonstrating better performance when using Schnorr than when using ECDSA. \textcolor{black}{Nevertheless, the TLS standard mandates the utilization of the ECDSA algorithm for digital signature creation, rendering the adoption of the Schnorr signature algorithm impractical now.}
%\vspace{-0.5cm}
\begin{table}[h!]
\caption{Modular multiplications in maTLS and mdTLS}\label{tab2}
\centering
\begin{tabular}{|C{50mm}|C{30mm}|C{30mm}|}
\hline
\textbf{Stages}  & \textbf{maTLS}      & \textbf{mdTLS}       \\
\hline
    \textbf{Certificate generation}        & 4,293N + 4,293 & 1,603N + 4,293   \\
    \textbf{Certificate verification}      & 1,792N + 1,792 & 897N + 1,792     \\
    \textbf{Security parameter blocks}     & 833N + 833     & 833N + 833        \\ \hhline{|===|}
    \textbf{Overall}     & \textbf{6,918N + 6,918}        & \textbf{3,333N + 6,918} \\
\hline
\end{tabular}
\end{table}

\vspace{-0.5cm}
\begin{figure}[ht!]
     \begin{subfigure}[b]{0.55\textwidth}
         \centering
         \includegraphics[width=\textwidth]{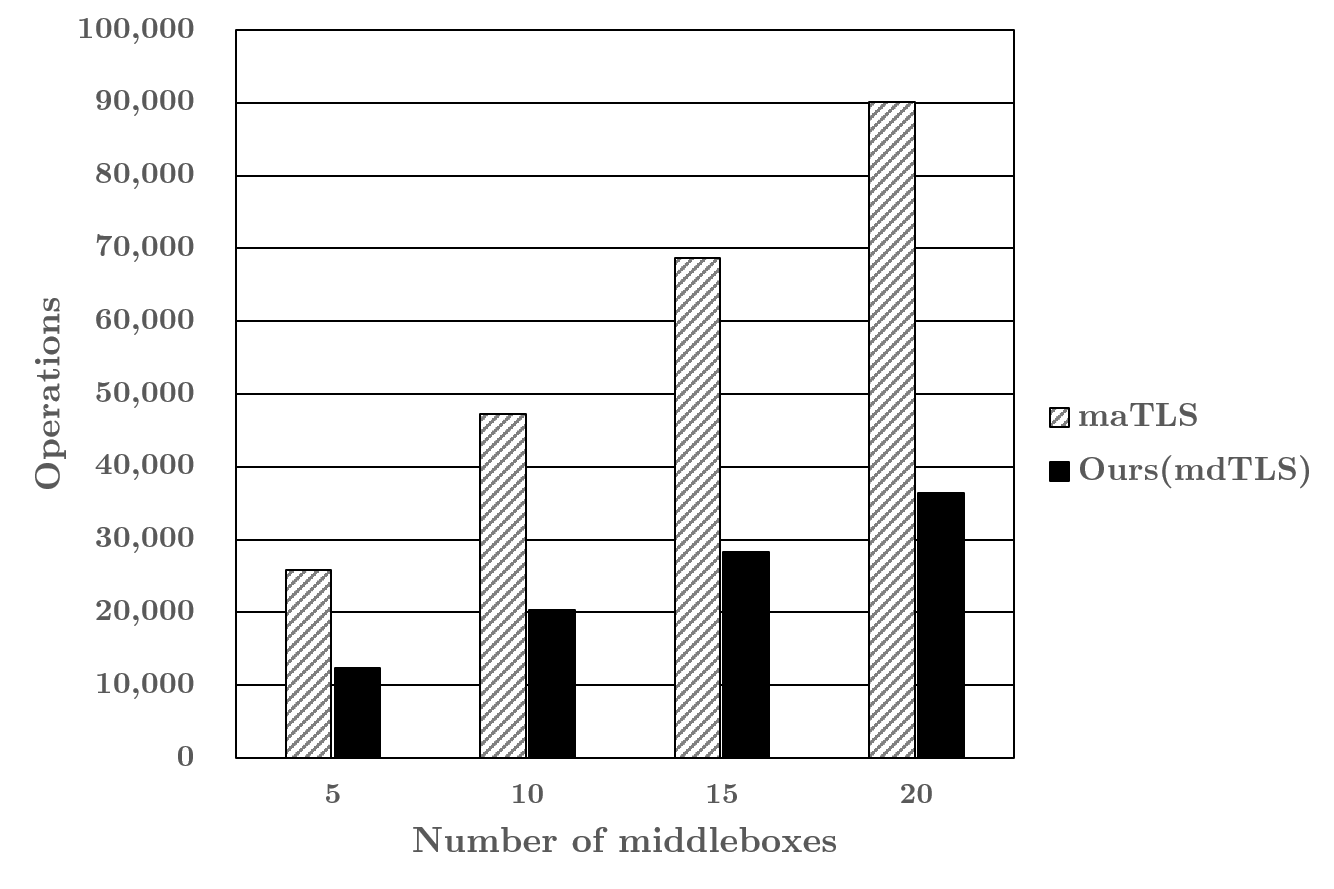}
         \caption{Certificate generation}
         \label{Certificate Generation}
     \end{subfigure}
     \hfill
     \begin{subfigure}[b]{0.55\textwidth}
         \centering
         \includegraphics[width=\textwidth]{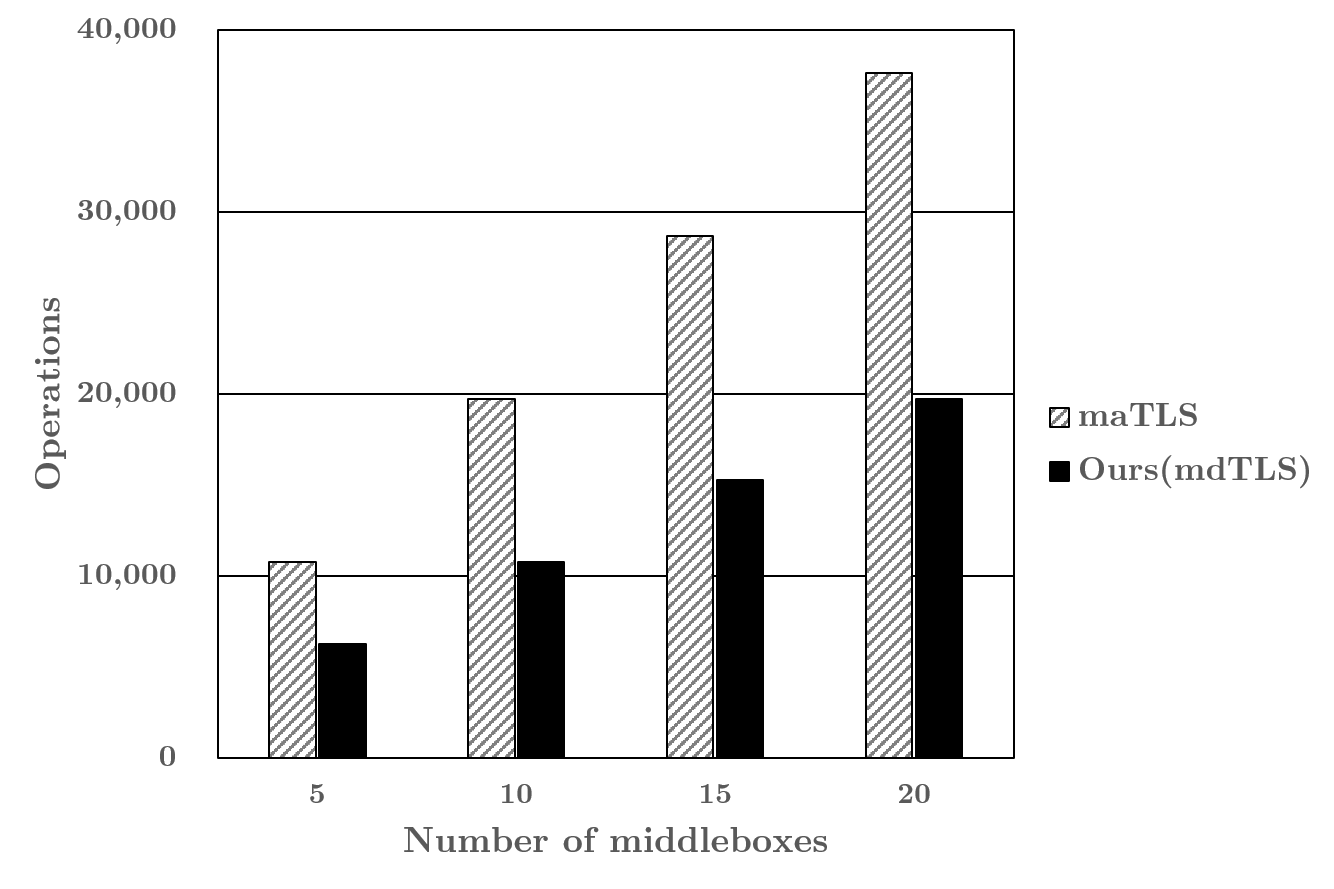}
         \caption{Certificate verification}
         \label{Certificate Verification}
     \end{subfigure}
     \hfill
     \begin{subfigure}[b]{0.55\textwidth}
         \centering
         \includegraphics[width=\textwidth]{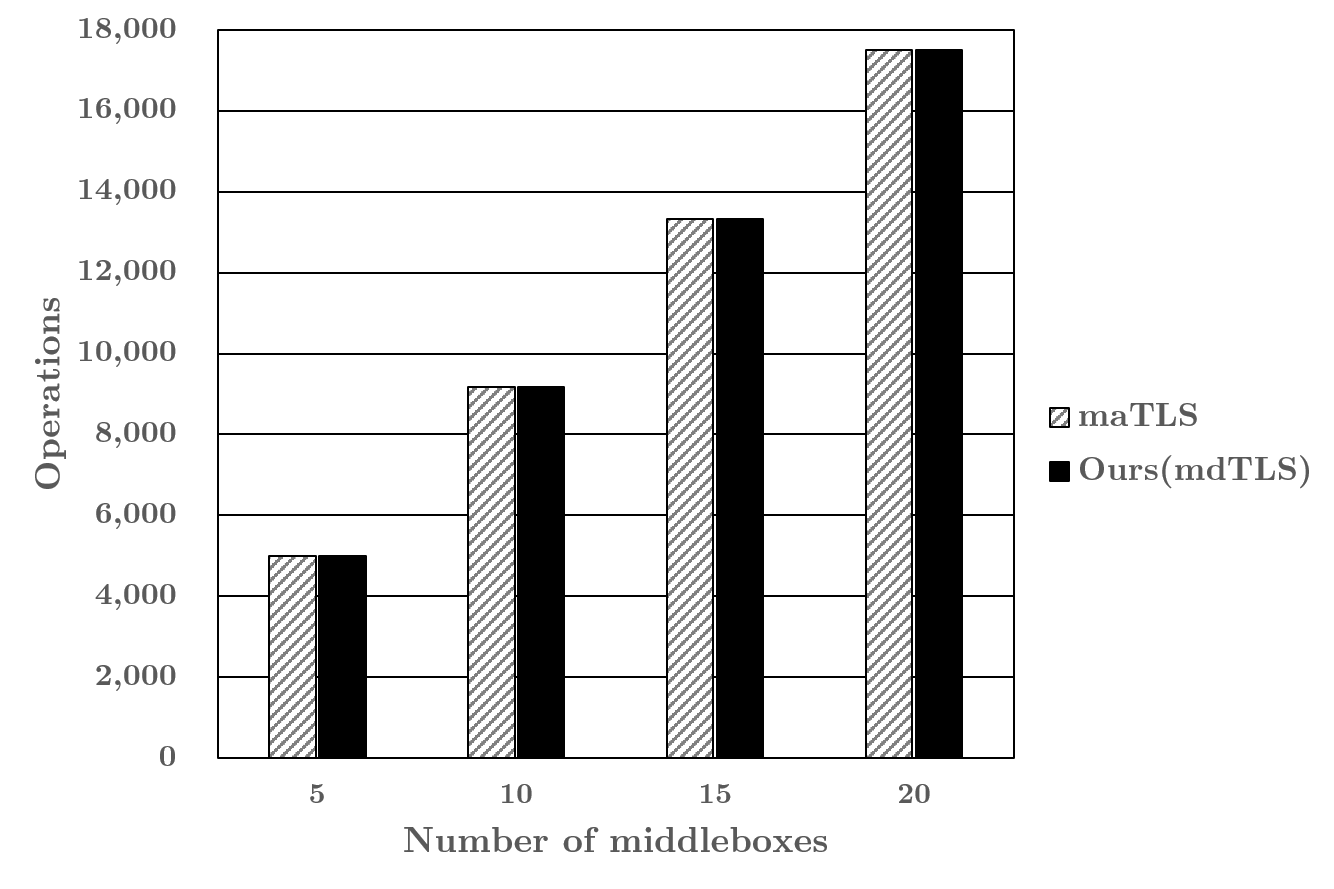}
         \caption{Security parameter blocks}
         \label{Security Parameter Blocks}
     \end{subfigure}
     \hfill
     \begin{subfigure}[b]{0.55\textwidth}
         \centering
         \includegraphics[width=\textwidth]{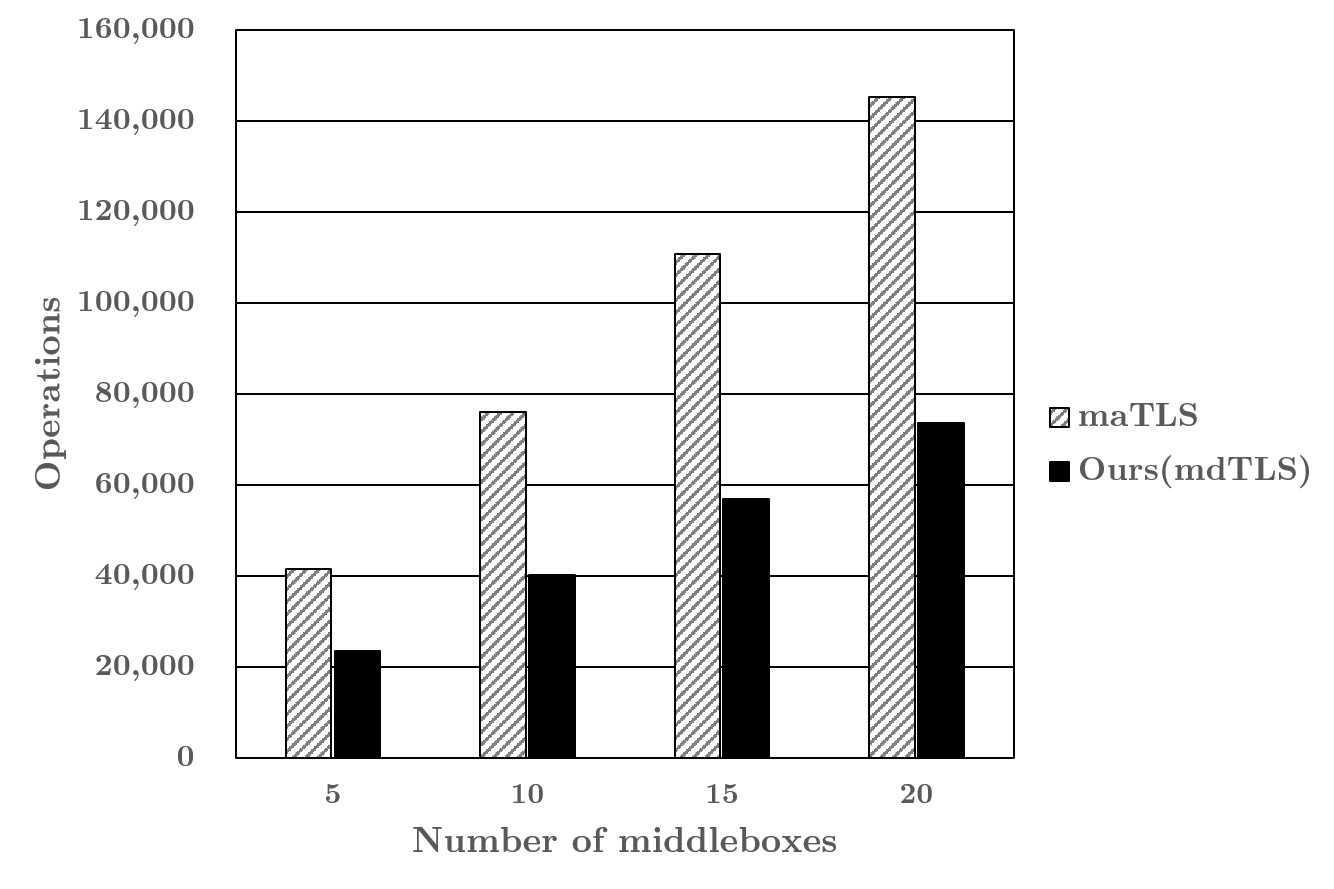}
         \caption{Overall performances}
         \label{Overall Performances}
     \end{subfigure}
        \caption{Performance of protocols when using Schnorr}
        \label{fig: graphs}
\end{figure}

\newpage
\section{Conclusion}
In this paper, we proposed a middlebox-delegated TLS protocol in which only middleboxes that have been permitted can participate in the network. To demonstrate the excellence of our proposed protocol, we verified our protocol from two aspects of view: performance and security. In the performance view, we calculated the number of computations in the protocol. We found out that the mdTLS reduces about 39\% of the computational complexity compared to maTLS. Also, we formally verified that our proposal achieved nine security lemmas: \textit{server/middlebox/data authentication}, \textit{path integrity}, \textit{path secrecy}, \textit{modification accountability}, \textit{verifiability}, \textit{strong-unforgeability}, and \textit{strong-identifiability}. Especially among them, the latter three security lemmas are newly defined for our protocol by extending existing concepts. The primary contribution of this work is to show that using the proxy signature scheme can enhance performance efficiency and maintain its security level.
\subsubsection{Acknowledgements}
This work was partly supported by Institute of Information communications Technology Planning Evaluation (IITP) grant funded by the Korea government (MSIT) (No.2018-0-00532, Development of High-Assurance (EAL6) Secure Microkernel, 100) and supported by Korea University.

\bibliographystyle{splncs04}
\bibliography{reference} % Entries are in the refs.bib file

\end{document}